\documentclass[twocolumn,trackchanges]{aastex701}
\usepackage{amsmath,amssymb}
\usepackage{graphicx}
\usepackage{color}
\usepackage{textgreek}

\DeclareRobustCommand{\revstrike}{\leavevmode\bgroup\ULdepth=-.55ex\ULset}

\newcommand{\Msun}{M_{\odot}}
\newcommand{\Ha}{\mathrm{H}\alpha}
\newcommand{\Hb}{\mathrm{H}\beta}
\newcommand{\Hg}{\mathrm{H}\gamma}

\newcommand{\cgsL}{\mathrm{erg~s^{-1}}}
\newcommand{\cgsF}{\mathrm{erg~s^{-1}~cm^{-2}}}

\newcommand\fullname{SDSS~J234728.73+242745.7}
\newcommand\srcname{J2347+2427}
\newcommand\tnsname{AT2020aczq}

\def\paperI{Paper~I}
\def\paperII{Paper~II}
\def\papers{Papers~I/II}
\def\papersa{Papers~I and II}

\makeatletter
\newcommand{\cmir}{\@ifnextchar\bgroup{\cmir@i}{\cmir@ii}}
\def\cmir@i#1{C_{\mathrm{MIR,#1}}}
\def\cmir@ii{C_{\mathrm{MIR}}}
\makeatother

\def\coefmir{\rho_{\mathrm{MIR}}}

\def\Fxray{F_{\mathrm{0.5-8~keV}}}

\def\Lbol{L_{\mathrm{bol}}}
\def\MBH{M_{\mathrm{BH}}}
\def\REdd{\lambda_\mathrm{Edd}}

\def\epochq{\textbf{\textit{S}$_{\mathrm{low}}$}}
\def\epochh{\textbf{\textit{S}$_{\mathrm{high}}$}}
\def\epochd{\textbf{\textit{S}$_{\mathrm{decl}}$}}
\def\epochr{\textbf{\textit{S}$_{\mathrm{rise}}$}}

\def\Fq{\left<F_{\mathrm{low}}\right>}
\def\Fh{\left<F_{\mathrm{high}}\right>}

\begin{document}

\title{Extreme Multiwavelength Variability in the Recurrent Intermediate-type Changing-look Seyfert Galaxy \fullname{}}

\correspondingauthor{Liming Dou; Jianguo Wang}
\email[show]{doulm@gzhu.edu.cn; wangjg@ynao.ac.cn}

\author[orcid=0009-0003-1518-6186]{Jiahua Wu}
\affiliation{Department of Astronomy, Guangzhou University, Guangzhou 510006, People’s Republic of China}
\affiliation{Department of Astronomy, Xiamen University, Xiamen, Fujian 361005, People’s Republic of China}
\email{jiahua@stu.xmu.edu.cn}

\author[orcid=0000-0002-4757-8622]{Liming Dou}
\affiliation{Department of Astronomy, Guangzhou University, Guangzhou 510006, People’s Republic of China}
\email{doulm@gzhu.edu.cn}

\author[orcid=0000-0003-4897-4106]{Yanli Ai}
\affiliation{College of Engineering Physics, Shenzhen Technology University, Shenzhen 518118, People’s Republic of China}
\email{aiyanli@sztu.edu.cn}

\author[orcid=0000-0003-4156-3793]{Jianguo Wang}
\affiliation{Key Laboratory for the Structure and Evolution of Celestial Objects, Yunnan Observatories, Kunming 650011, People’s Republic of China}
\email{wangjg@ynao.ac.cn}

\author[orcid=0000-0002-1517-6792]{Tinggui Wang}
\affiliation{Department of Astronomy, University of Science and Technology of China, Hefei, 230026, China}
\affiliation{School of Astronomy and Space Sciences, University of Science and Technology of China, Hefei, 230026, China}
\affiliation{Department of Physics and Astronomy, College of Physics, Guizhou University, Guiyang 550025, People's Republic of China}
\email[]{twang@ustc.edu.cn}

\author[orcid=0000-0003-4200-9954]{Huaqing Cheng}
\affiliation{National Astronomical Observatories, Chinese Academy of Sciences, Beijing 100101, China}
\email{hqcheng@nao.cas.cn}

\author[orcid=0000-0002-2006-1615]{Chichuan Jin}
\affiliation{National Astronomical Observatories, Chinese Academy of Sciences, Beijing 100101, China}
\affiliation{School of Astronomy and Space Science, University of Chinese Academy of Sciences, 19A Yuquan Road, Beijing 100049, China}
\affiliation{Institute for Frontier in Astronomy and Astrophysics, Beijing Normal University, Beijing 102206, China}
\email{ccjin@nao.cas.cn}

\author[orcid=0000-0001-6947-5846]{Luis C. Ho}
\affiliation{Kavli Institute for Astronomy and Astrophysics, Peking University, Beijing 100871, China}
\affiliation{Department of Astronomy, School of Physics, Peking University, Beijing 100871, China}
\email{lho.pku@gmail.com}

\author[orcid=0009-0005-4743-5364]{Qiming Wu}
\affiliation{Department of Astronomy, Guangzhou University, Guangzhou 510006, People’s Republic of China}
\email{qimingwu@e.gzhu.edu.cn}

\author[orcid=0000-0002-7329-3209]{Kovi Rose}
\affiliation{Sydney Institute for Astronomy, School of Physics, The University of Sydney, NSW 2006, Australia}
\affiliation{Australia Telescope National Facility, CSIRO, Space \& Astronomy, PO Box 76, Epping, NSW 1710, Australia}
\email{kovi.rose@sydney.edu.au}

\author[orcid=0000-0002-4622-796X]{Roberto Soria}
\affiliation{INAF--Osservatorio Astrofisico di Torino, Strada Osservatorio 20, I-10025 Pino Torinese, Italy}
\affiliation{Sydney Institute for Astronomy, School of Physics A28, The University of Sydney, Sydney NSW 2006, Australia}
\email{roberto.soria@inaf.it}

\author[orcid=0000-0003-0860-440X]{Alexis Coleiro}
\affiliation{Université Paris Cité, CNRS, Astroparticule et Cosmologie, F75013 Paris, France}
\email{alexis.coleiro@u-paris.fr}

\author[orcid=0000-0002-5929-0968]{Junhui Fan}
\affiliation{Department of Astronomy, Guangzhou University, Guangzhou 510006, People’s Republic of China}
\email{fjh@gzhu.edu.cn}

\author[orcid=0009-0007-6104-337X]{Yuan Liu}
\affiliation{National Astronomical Observatories, Chinese Academy of Sciences, Beijing 100101, China}
\email{liuyuan@bao.ac.cn}

\author[orcid=0000-0002-4562-7179]{Dongyue Li}
\affiliation{National Astronomical Observatories, Chinese Academy of Sciences, Beijing 100101, China}
\email{dyli@nao.cas.cn}

\author[orcid=0000-0002-8636-8601]{Zhen Zhang}
\affiliation{National Astronomical Observatories, Chinese Academy of Sciences, Beijing 100101, China}
\email{zhangzhen@nao.cas.cn}

\author[orcid=0000-0002-7397-811X]{Yunfei Xu}
\affiliation{National Astronomical Observatories, Chinese Academy of Sciences, Beijing 100101, China}
\email{xuyf@nao.cas.cn}

\begin{abstract}
We report the discovery of the Seyfert galaxy J234728.73+242745.7 as a recurrent intermediate-type changing-look active galactic nucleus (CLAGN) with extreme optical/UV variability.
Optical spectra spanning nearly three decades reveal Type~1.0$\rightarrow$1.5$\rightarrow$1.2 transitions, accompanied by the disappearance and reappearance of broad \ion{He}{2}~$\lambda4686$ and a broad-line Balmer decrement evolving from 4.5 to 7.2 and then 2.4.
Multi-wavelength monitoring reveals coordinated variability from X-rays to the mid-infrared (MIR).
Spectral energy distribution decomposition yields a host-corrected $g$-band high-to-low-state flux ratio of $5.4\pm0.4$, exceeding the Swift/XRT ratio of $2.7\pm0.8$ across the changing-look transition.
This discrepancy may arise from reduced dust extinction, as correcting for the low-state continuum reddening ($A_V^{\mathrm{cont}}=0.66\pm0.23$~mag) brings the optical/UV ratios close to the X-ray value, although the correction is model-dependent and intrinsically weaker X-ray variability may also contribute.
However, X-ray spectra show no compelling evidence for intrinsic gas absorption or its variability ($N_{\mathrm{H}}^{\mathrm{intr}}\lesssim5\times10^{20}$~cm$^{-2}$).
In contrast, EP/WXT monitoring during the recent high state reveals order-of-magnitude X-ray fluctuations on monthly-to-sub-year timescales, exceeding contemporaneous optical variations.
Meanwhile, the radio flux density remains stable at $\sim2$~mJy across all ASKAP observations since mid-2020.
Together with optical/UV and MIR variability and the broad \ion{He}{2} transition, these results favor intrinsic accretion-state changes as the primary driver, while allowing a secondary contribution from variable obscuration.
The source displays a loop-like MIR color--magnitude evolution, likely driven by the delayed response of an extended torus. This recurrent CLAGN offers a laboratory for probing the dynamic response of the broad-line region (BLR) and circumnuclear dust to accretion-state transitions.
\end{abstract}


\section{Introduction}
\label{sec:intro}

Active galactic nuclei (AGN; \citealt{Seyfert1943, Schmidt1963}) are powered by accretion onto supermassive black holes (SMBHs). In the standard paradigm, an accretion disk produces the optical/ultraviolet (UV) continuum, a hot corona generates X-rays, and a circumnuclear dust torus reprocesses the central radiation into the infrared (IR).
AGN are traditionally classified as Type~1 or Type~2 based on their optical spectra.
Within the orientation-based unification model, Type~1 AGN show both broad and narrow emission lines from the broad-line region (BLR) and narrow-line region (NLR), respectively, whereas in Type~2 AGN, the BLR and nuclear continuum are obscured by the dust torus, leaving only narrow emission lines visible \citep{Antonucci1993}.

Although all Type~1 AGN exhibit broad $\Ha$ and/or $\Hb$ emission lines, the visibility and relative strength of these broad components span a continuous range, giving rise to the spectral subclasses Types~1.0, 1.2, 1.5, 1.8, and 1.9.
These intermediate subtypes were traditionally distinguished through visual inspection \citep{Osterbrock1977}, and later quantified using empirical flux ratios involving broad $\Hb$ and [O~III]~$\lambda$5007 (hereafter $\Hb$/[\ion{O}{3}]; \citealt{Whittle1992,Winkler1992,Maiolino1995}).
These subtypes are often interpreted as an obscuration sequence, representing progressively increasing levels of nuclear obscuration toward Type~2 AGN \citep{Maiolino1995}. More recent studies suggest that, at least among Type~1 subtypes, the observed spectral sequence can be explained by an accretion sequence, reflecting different accretion states rather than different levels of obscuration \citep{Burtscher2016, Tan2026}.

Contrary to this static picture, an increasing number of AGN have been found to undergo dramatic spectral transitions on human timescales. The hallmark of these changing-look AGN (CLAGN; \citealt{LaMassa2015, MacLeod2016, MacLeod2019, Ricci2023}) is the complete emergence or disappearance of broad $\Ha$ and/or $\Hb$ emission lines, accompanied by large continuum variations, driving transitions between Type~1.0--1.8 and Type~1.9/2.
Such objects are commonly referred to as turn-on/off CLAGN.

Recently, the changing-look concept has been extended to transitions among intermediate subtypes \citep[e.g.,][]{Hon2020, Senarath2021}, even when broad emission lines never disappear completely, motivated by the growing number of sources showing subtype (or $\Hb$/[\ion{O}{3}]) variations on year-to-decade timescales \citep{Amrutha2026}.

Changing-look behavior has now been identified over a broad range of AGN luminosities and across diverse AGN subclasses, from nearby Seyfert galaxies to luminous quasars and LINERs \citep[e.g.,][]{Frederick2019}.
Even blazars, whose optical spectra can exhibit a wide range of emission-line properties or featureless continua \citep{ChenGH2026a}, have been observed to undergo dramatic spectral-state transitions (e.g., Ton~599; \citealt{Chen2025}).
More broadly defined changing-state phenomena have also been reported beyond optical emission-line changes, such as the emergence of a radio jet in the narrow-line Seyfert~1 galaxy J1105+1452 \citep{Dou2026, Komossa2026}.
These observations suggest that changing-state behavior is not restricted to a particular AGN subclass or luminosity regime, although the underlying physical mechanisms may differ among individual systems.

The physical origin of changing-look behavior remains an active topic of investigation. Growing observational evidence suggests that, for many optical CLAGN, the dramatic continuum and broad-line changes are primarily driven by intrinsic variations in the central engine rather than by variable line-of-sight (LoS) obscuration \citep[e.g.,][]{LaMassa2015, MacLeod2016, Sheng2017, Ricci2023}. Such variations may arise from substantial changes in the accretion rate and/or the structure of the inner accretion flow, including disk instabilities and transitions between different accretion states \citep[e.g.,][]{Noda2018, Sheng2020, Wang2024, Dong2025, LiCao2026}. These changes can substantially modify the ionizing continuum and consequently alter the ionization state of the BLR, leading to the weakening, disappearance, or reappearance of broad emission lines.
Recent studies of recurrent and rapidly transitioning CLAGN further suggest that such changes in the inner accretion flow can occur on timescales shorter than the viscous timescale expected for a standard thin disk \citep[e.g.,][]{Wang2024, Wang2025, Dong2025, Dong2026, LiCao2026, ChenGH2026b}. Other mechanisms may also contribute in individual sources. Transient accretion events, such as tidal disruption events (TDEs), can produce dramatic and often transient luminosity changes \citep[e.g.,][]{Merloni2015, Li2022, Li2024a, Li2024b, Son2022b}, while variable obscuration may account for a subset of changing-look events \citep[e.g.,][]{Gilli2000, Ricci2023}. X-ray observations provide a particularly powerful diagnostic because they directly probe the innermost disk-corona emission while also constraining variable gas absorption along the LoS, thereby helping distinguish intrinsic accretion changes from obscuration-driven variability \citep[e.g.,][]{Ai2020, Ricci2023}.

Mid-infrared (MIR) observations provide an important complementary probe of AGN variability.
The MIR emission arises primarily from hot dust heated by the accretion disk.
Its variability directly reflects dust responses, producing a ``dust echo'' that encodes both the dust temperature evolution and the geometry of the obscuring structure.
The \textit{Wide-field Infrared Survey Explorer} (WISE; \citealt{Wright2010}) enabled decade-long, all-sky MIR monitoring, revealing stronger MIR variability in AGN than in inactive galaxies due to dust reprocessing. The MIR color criterion $\cmir = W1-W2 \gtrsim 0.8$ efficiently separates AGN from stars and normal galaxies \citep{Assef2013}, where $W1$ and $W2$ denote the 3.4 and 4.6~$\mu$m WISE bands.

In \citet[][hereafter \paperI{}]{nlsy1_mcv} and \citet[][hereafter \paperII{}]{sy1_mcv}, we systematically analyzed MIR color variations (MCVs) in broad-line AGN samples drawn from \citet{Paliya2024}.
We showed that a variable-temperature hot dust component diluted by invariant host starlight can reproduce different MCV trends; the host fraction determines whether the MCV is redder-when-brighter (RWB; host dominated), bluer-when-brighter (BWB; dominated by AGN-heated dust), or weak/no MCV (comparable contributions).
This framework further predicts that RWB sources should exhibit bluer minimum MIR colors $\cmir{min}$ and larger MCV amplitudes $\Delta\cmir$ than BWB sources (see Figures~5 and 6 of \papersa{}, respectively).
Interestingly, previously known turn-on/off CLAGN (e.g., \citealt{Frederick2019, Lyu2022}) predominantly overlap with the RWB population in the $\cmir{min}$--$\Delta\cmir$ plane, indicating that the changing-look phenomenon is characterized by distinctive MIR color evolution.

\fullname{} (hereafter \srcname{}), classified as a radio-quiet broad-line Seyfert~1 with a redshift of $z=0.1571$ by \citet{Paliya2024}, attracted our attention during the analysis of \paperII{} because of its unusual MCV behavior.
Although formally classified as a weak MCV source, its MIR color–magnitude diagram exhibits a prominent hysteresis loop rather than the reversible single-path evolution predicted by the host-dilution framework under the assumption of single-temperature, synchronous blackbody emission (see Figure~6 in \paperI{}), indicating a more complex physical process. In addition, it exhibits exceptionally large MIR variability, with amplitudes ($\sigma_m[W1]=0.45$~mag) exceeding those of $\sim$99.7\% of the Seyfert~1s with significant MIR variability in \paperII{}.

The source also displays rapid and strong optical variability: it brightened by $\sim 1$~mag in the $G$ band within one year around 2020 December (MJD~59200.08), triggering a Gaia Science Alert (\tnsname{}; \citealt{AT2020aczqTNSTR}).
This alert prompted several months of Swift/XRT monitoring.
Our subsequent spectroscopic monitoring reveals that the source is a recurrent intermediate-type CLAGN, exhibiting a Type~1.0$\rightarrow$1.5$\rightarrow$1.2 transition over the past three decades. \srcname{} provides an excellent opportunity to investigate the physical mechanisms underlying the changing-look phenomenon, as well as the response of the accretion flow, BLR, and circumnuclear dust to dramatic changes in the central engine.

In this manuscript, we present a comprehensive multi-wavelength analysis of \srcname{} spanning X-ray to MIR.
We adopt a flat $\Lambda$CDM cosmology with $H_0 = 67.4~\mathrm{km~s^{-1}~Mpc^{-1}}$, $\Omega_{\mathrm{m}} = 0.315$, and $\Omega_{\Lambda} = 0.685$ \citep{Planck2018}. For \srcname{} at $z=0.1571$, this corresponds to a luminosity distance of $D_{\mathrm{L}}\simeq778~\mathrm{Mpc}$.

\section{Observations and Data Reduction}
\label{sec:obs}

\subsection{Optical and Infrared Photometric Monitoring}
\label{sec:lc}
\begin{figure}[hbt!]
\centering
\includegraphics[width=\columnwidth]{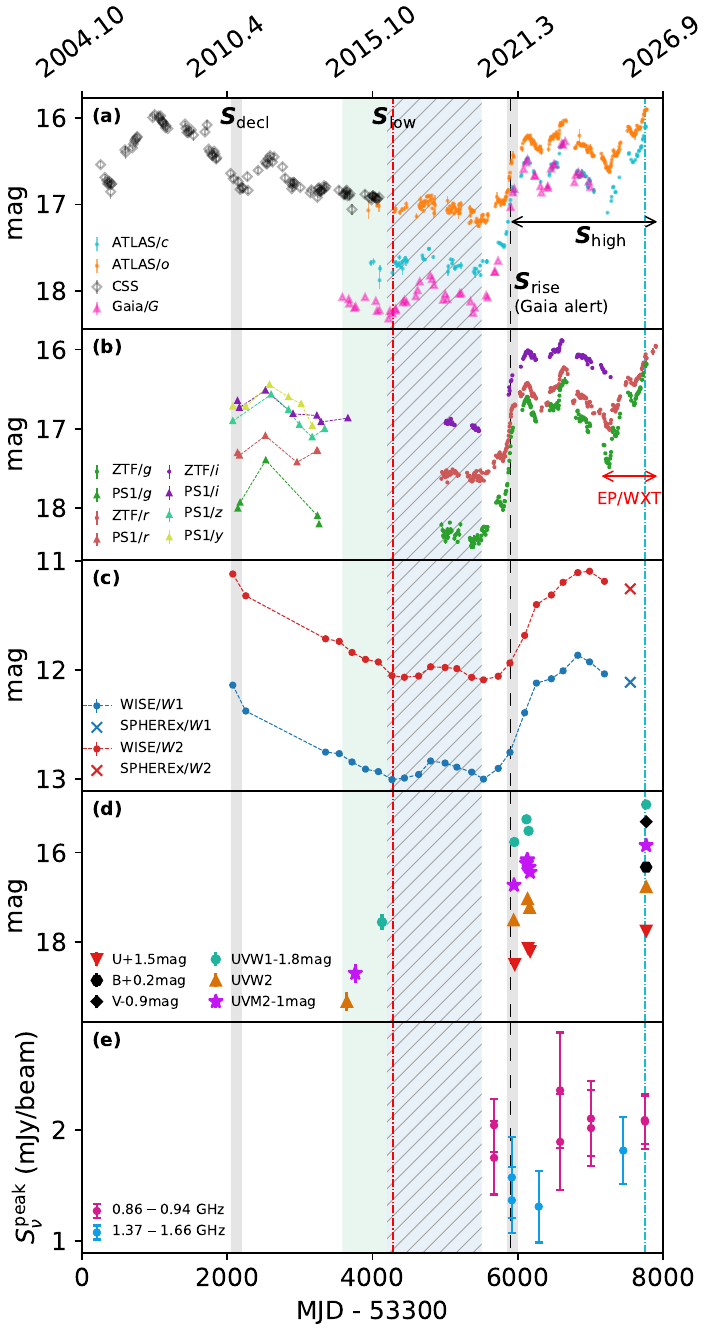}
\caption{
Multi-wavelength light curves of \srcname{} across the optical/UV, MIR, and radio bands.
The vertical black dashed line marks the Gaia Alert date.
The green (620~days) and diagonally hatched blue (1300~days) shaded regions together mark the extended low-flux period (\epochq{}): the optical emission becomes quiescent first in the green region, while the WISE flux continues to decline, with both settling into quiescence only in the blue region.
The two gray shaded regions denote the 150-day SED windows for the declining (\epochd{}) and rising (\epochr{}) phases (Section~\ref{sec:sed}).
Vertical red and cyan dashed lines indicate the spectroscopic observation epochs from the SDSS and Lijiang 2.4~m telescope, respectively.
\label{fig:lc}}
\end{figure}

Optical light curves (LCs) were assembled from multiple surveys.
We retrieved Gaia $G$-band photometry from the Gaia Science Alerts archive \citep{Gaia2023}, which provides multi-year monitoring covering the pre- and post-alert epochs.
Early-time coverage is provided by the Catalina Sky Survey (CSS; \citealt{Drake2009}), which monitored the source in an unfiltered $V$-band analogue before 2016 January. We restricted the CSS photometry to \texttt{blend} = 0 to avoid contamination from neighboring sources.
The Asteroid Terrestrial-impact Last Alert System (ATLAS; \citealt{Tonry2018}) provides $c$ and $o$-band forced photometry (\citealt{Shingles2021}); we selected measurements with $\chi^2/N < 5$ and above the $5\sigma$ detection limit.
Together, the Gaia, CSS, and ATLAS data provide continuous optical coverage from 2005 July to 2026 February, and are presented in Figure~\ref{fig:lc}a.

High-cadence multi-band photometry was obtained from the Zwicky Transient Facility (ZTF; \citealt{Masci2019}) in the $gri$ bands, retaining only high-quality detections (\texttt{catflags} = 0).
We also retrieved $grizy$ aperture photometry from Pan-STARRS1 (PS1; \citealt{Flewelling2020}), requiring \texttt{airmass} $\leq 1.4$ and a reliable aperture filling factor \texttt{apFillFactor} $> 0.8$.
These data are shown in Figure~\ref{fig:lc}b.

For each optical survey, outliers were removed by iteratively clipping points deviating by $>3\sigma$ from a 15-day sliding-window median.
The cleaned LCs were binned into 5-day intervals to reduce scatter while preserving the intrinsic variability.
All optical magnitudes were corrected for Galactic extinction using the dust maps of \citet{Schlafly2011} and the Milky Way extinction curve of \citet{Cardelli1989} with $R^{\mathrm{MW}}_V = 3.1$.

For the MIR, we used multi-epoch photometry from the AllWISE and NEOWISE missions \citep{Wright2010}, following the data reduction procedure of \paperII{} to construct $W1$ and $W2$ LCs spanning 2010--2024 at a typical six-month cadence.
We also included data from the \textit{Spectro-Photometer for the History of the Universe, Epoch of Reionization and Ices Explorer} (SPHEREx; launched 2025 March; \citealt{Bock2026}), which provides all-sky 0.75–5~$\mu$m spectroscopy. From the SPHEREx spectra, we synthesized $W1$ and $W2$ fluxes and transformed them to the WISE photometric system.
The MIR LCs are shown in Figure~\ref{fig:lc}c.

The CSS, ATLAS, and Gaia LCs together span nearly two decades, revealing a broad transition from an early high state to a prolonged low-flux phase. After several years of quiescence, the source brightened rapidly around the Gaia alert (MJD 59200.08; black dashed line) and entered a renewed high state lasting $\sim$4.5~yr (2021.8--2026.2; MJD 59427--61069).
The MIR emission broadly follows the optical evolution, fading into the low state and brightening after the Gaia alert.

We defined a low state \epochq{} and a recent high state \epochh{}.
\epochh{} corresponds to the period from the Gaia alert to the present.
\epochq{} spans $\sim5$~yr, during which the optical emission stays at a low flux level. In Figure~\ref{fig:lc} we mark \epochq{} with two shaded intervals: a green region of 620~days and a blue region of 1300~days with diagonal hatching. The blue hatched region indicates the stage where both the optical and MIR emission have fully entered the quiescent state.
During the first $\sim1.5$~yr of \epochq{} (i.e., green shading), the WISE fluxes are still declining significantly and continuously.

\subsection{Archival SDSS and GALEX Photometry}
We also searched the SDSS imaging archive and the Galaxy Evolution Explorer (GALEX) archive for complementary early-time photometry.\footnote{The SDSS and GALEX magnitudes quoted below are also corrected for Galactic extinction.}

The primary SDSS detection (ObjID~1237678579827474520) was obtained on 2008 October~1 (MJD~54740.394) and has \texttt{clean}=1. We adopt the \texttt{modelMag} photometry, obtaining $u=16.834\pm0.009$, $g=16.744\pm0.004$, $r=16.513\pm0.004$, $i=16.037\pm0.004$, and $z=16.157\pm0.008$~mag. An earlier secondary detection on 2004 September~17 (MJD~53265.235; ObjID~1237666244691886261) has \texttt{clean}=0; we therefore do not use it for quantitative variability measurements.
GALEX detected a counterpart within $1.1\arcsec$ of the optical position in the AIS field \texttt{AIS\_145\_1\_71}, observed on 2006 October~26--29 (MJD~54034.416--54037.361), with a total exposure of 204~s in each UV band. The magnitudes are $NUV=17.86\pm0.04$ and $FUV=18.21\pm0.07$~mag, and both artifact flags are zero. The quoted values are AB magnitudes.

Although differences in filter response limit a direct comparison, the 2008 SDSS photometric measurements are broadly consistent with a relatively high optical flux state before \epochq{}. Similarly, the 2006 GALEX measurements indicate a relatively high early UV flux level compared with the subsequent Swift/UVOT observations (see Section~\ref{sec:data_uvot}).

\subsection{Spectroscopic Observations}
Archival spectra include a low-resolution objective-prism spectrum from the Hamburg Quasar Survey (HS~2344+2411; 1997 August), obtained with the 80~cm Hamburg Schmidt telescope at Calar Alto Observatory as part of a wide-field slitless spectroscopic survey \citep{Hagen1995}, and an SDSS/eBOSS spectrum (2016 July, MJD 57580) obtained during the low state \epochq{}.

We acquired two recent high-state spectra (2026 Jan.~10 and Feb.~7, hereafter LJ110 and LJ207) with the YFOSC spectrograph on the Lijiang 2.4~m telescope (LJT). Both cover 3858--7330\,\AA\ with a resolution of $R \approx 280$ and were reduced using standard \texttt{IRAF} procedures.

\begin{figure*}[hbt!]
\centering
\plotone{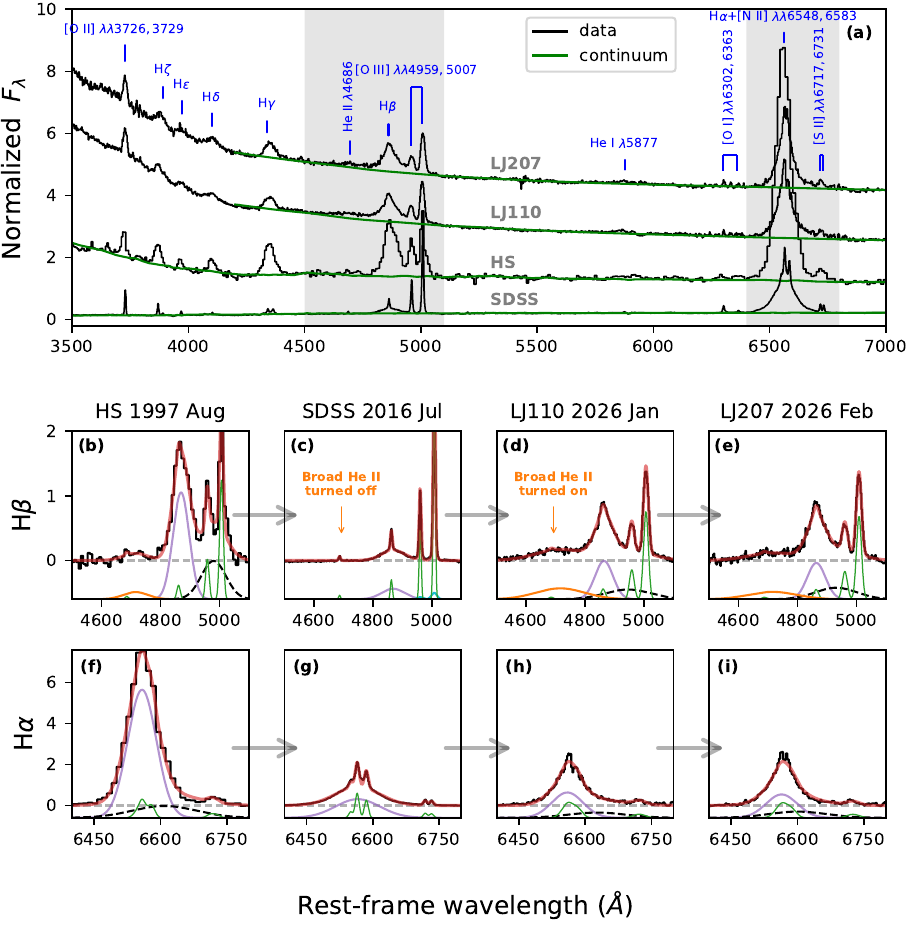}
\caption{
\textit{Panel} (a): Optical spectra of \srcname{}: HS, SDSS, and the two LJT spectra, all scaled to the [\ion{O}{3}]~$\lambda5007$ core and vertically offset for clarity. The shaded regions mark the $\Hb$ (4500--5100\,\AA) and $\Ha$ (6400--6800\,\AA) regions.
\textit{Panels} (b)--(e): Zoom-in of the continuum-subtracted $\Hb$ region for each spectrum, shown in chronological order from left to right. The spectra include broad \ion{He}{2}~$\lambda4686$ (orange), broad Balmer lines (purple), and narrow lines (green). The [\ion{O}{3}] wing in the SDSS spectrum is shown in cyan, while in the other spectra the [\ion{O}{3}] wing and excess emission are jointly modeled with an additional broad Gaussian (dashed black).
\textit{Panels} (f)--(i): same as (b)--(e), but for the $\Ha$ region.
\label{fig:spec}}
\end{figure*}

The four spectra are shown in Figure~\ref{fig:spec}a, with their corresponding epochs marked in Figure~\ref{fig:lc}. All spectra were corrected for Galactic extinction and flux-calibrated by scaling the [\ion{O}{3}] $\lambda5007$ to the SDSS value during the emission-line modeling described in Section~\ref{sec:opt_spectra}.

\subsection{Swift Observatory}
\subsubsection{X-ray Telescope}
\srcname{} was observed multiple times by the Neil Gehrels \textit{Swift} Observatory \citep{Gehrels2004}. We retrieved all archival X-ray Telescope (XRT; \citealt{Burrows2005}) observations with exposure times $>$ 150~s, yielding 18 usable epochs. Three observations were obtained during \epochq{}, while the remaining fifteen were taken after the Gaia alert, including a 2.7~ks observation that we requested in 2026 January. The individual exposure times range from 0.157 to 2.7~ks.
All XRT observations were performed in photon counting mode \citep{Hill2004}. The data were reduced using \texttt{HEASOFT} (v6.35) and the \texttt{xrtpipeline} task. Source events were extracted from a circular aperture with a radius of $50\arcsec$ centered on the optical position, while the background was estimated from a nearby source-free region.

\subsubsection{UV--Optical Telescope}
\label{sec:data_uvot}
During the XRT observations, the Swift UV--Optical Telescope (UVOT; \citealt{Roming2005}) obtained simultaneous photometry in the optical and UV. Observations were taken in six filters: $V$ (5468\,\AA), $B$ (4392\,\AA), $U$ (3465\,\AA), $UVW1$ (2600\,\AA), $UVM2$ (2246\,\AA), and $UVW2$ (1928\,\AA). The full set of six filters was used only for our requested 2026 Jan. observation; all other epochs provided a subset.

Source counts were extracted within a $10\arcsec$ radius aperture centered on the target, with the background estimated from a nearby $20\arcsec$ source-free region. After background subtraction, AB magnitudes were derived with \texttt{uvotsource} using the most recent calibrations \citep{Poole2008, Breeveld2010} and corrected for Galactic extinction. Only measurements with detection significance $>3\sigma$ were retained.
The UVOT LCs are shown in Figure~\ref{fig:lc}d and are consistent with the optical variability.

\subsection{Einstein Probe}

\subsubsection{Wide-field X-ray Telescope}
\label{sec:wxtdata}

The Wide-field X-ray Telescope (WXT) on board the Einstein Probe (EP; \citealt{Yuan2022}) is a wide-field X-ray monitor employing novel lobster-eye micro-pore optics, providing a field of view (FOV) of $\sim3600$~deg$^2$. Operating in the 0.5--4~keV band, WXT reaches a sensitivity of $(2$--$3)\times10^{-11}$~erg~cm$^{-2}$~s$^{-1}$ for a typical 1~ks exposure. The energy resolution is $\sim170$~eV at 1~keV, and the angular resolution is $\sim5^\prime$ (FWHM).

The position of \srcname\ was covered repeatedly by EP/WXT between 2024~July and 2026~July. Photon events were extracted from a circular aperture with a radius of $3^\prime$. Because the source is too faint to be significantly detected in individual WXT snapshots, the observations were grouped into 14-day bins to improve the signal-to-noise ratio.

To assess whether the apparent WXT variability could be contaminated by nearby X-ray sources, we visually inspected the most recent 2.7~ks Swift/XRT image, which provides substantially higher angular resolution than WXT. No significant X-ray source other than \srcname{} is visible within the WXT aperture. We further cross-matched the region with the Living Swift XRT Point Source (LSXPS; \citealt{Evans2023}) catalog and found no additional X-ray sources within the aperture.

\subsubsection{Follow-up X-ray Telescope}
The Follow-up X-ray Telescope (FXT; \citealt{Chen2021}) on board EP is a Wolter-I telescope operating in the 0.3--10~keV band. EP/FXT consists of two identical modules, FXT-A and FXT-B.
FXT observed \srcname{} simultaneously with both modules on 2026~July~5 in the Full Frame (FF) mode, with an effective exposure of 2.4~ks for each module. The data were reduced using the Follow-up X-ray Telescope Data Analysis Software (\texttt{FXTDAS}\footnote{\url{https://epfxt.ihep.ac.cn/analysis}}, v1.30). Source events were extracted from a circular aperture with a radius of $100\arcsec$ centered on the target, while the background was estimated from a nearby source-free circular region.

\subsection{ROSAT Archival Data}

\srcname{} was detected in the ROSAT All-Sky Survey (RASS; \citealt{Voges1999}) from sequence \texttt{rs931361n00}.
We retrieved the archival Position Sensitive Proportional Counter (PSPC) event data and extracted source photons from a circular aperture of radius $510\farcs6$ centered on the source position. The background was estimated from a nearby source-free region.
Inspection of the archival event data confirmed that the source was observed on 1990 July~11--13 and December~20--21.
Because the \texttt{ONTIME} keyword records the total accepted time rather than the local exposure in RASS scanning-mode data, we adopted an exposure time of 694~s from the exposure map.

\subsection{ASKAP Radio Observations}
We collected radio observations of \srcname\ from the \textit{Australian SKA Pathfinder} (ASKAP; \citealt{Hotan2021}) telescope. The source is detected as an unresolved radio source in multiple ASKAP observations from the Rapid ASKAP Continuum Survey (RACS; \citealt{McConnell2020, Hale2021}) and the First Large Absorption Survey in H\,{\sc i} (FLASH; \citealt{Allison2022, Yoon2025}), spanning MJD 58972--61050. The radio continuum images show a compact source at the position of \srcname, with no obvious extended emission. We measure the peak flux density, $S^{\rm peak}_{\nu}$, at the source position from each ASKAP image, yielding values of approximately 1.3--2.4~mJy across the central frequencies of 0.85--1.65~GHz (see Figure~\ref{fig:lc}e for the ASKAP LC). These ASKAP images are averaged across the full 15~min or 2~hr observation, for RACS and FLASH respectively, and a bandwidth of 288~MHz.

\section{Analysis and Results}

\subsection{Optical Spectral Analysis}
\label{sec:opt_spectra}
\subsubsection{Continuum and Emission-line Fitting}
We use the SDSS and LJT spectra to study the spectral evolution of \srcname\ before and after the 2020 outburst, while the HS spectrum provides an additional view of its earlier state. We first perform a spectral decomposition of the highest-resolution SDSS spectrum and then use the resulting narrow lines to constrain other spectra.

We fit the SDSS spectrum over the rest-frame wavelength range of 3500--8000~\AA\ using the \textit{Bayesian AGN Decomposition Analysis for SDSS Spectra} (\texttt{BADASS}\footnote{\url{https://github.com/remingtonsexton/BADASS3}}; \citealt{Sexton2021}).
The continuum is modeled as the sum of host starlight and AGN emission. The host component is represented by four single stellar population templates with ages of 0.1, 1, 5, and 10~Gyr from the EMILES library \citep{Vazdekis2016}, while the AGN component consists of a smoothly broken power law plus the optical \ion{Fe}{2} template of \citet{VeronCetty2004}.

For the emission lines, we model the $\Hb$ (4500--5100~\AA) and $\Ha$ (6400--6800~\AA) regions. The $\Hb$ region includes narrow \ion{He}{2}~$\lambda4686$, [\ion{O}{3}]~$\lambda\lambda4959,5007$, and $\Hb$, together with broad $\Hb$ and \ion{He}{2}~$\lambda4686$. We also consider a broad [\ion{O}{3}] wing component. The $\Ha$ region includes narrow $\Ha$, [\ion{N}{2}]~$\lambda\lambda6548,6583$, and [\ion{S}{2}]~$\lambda\lambda6716,6731$, together with broad $\Ha$. Each component is modeled with a Gaussian profile.
In the $\Hb$ region, the widths and velocity offsets of all narrow emission lines are tied to those of the [\ion{O}{3}]~$\lambda5007$ core, while those in the $\Ha$ region are tied to [\ion{S}{2}]~$\lambda6731$. The flux ratios of the [\ion{O}{3}] and [\ion{N}{2}] doublets are fixed to 3:1.

For the HS spectrum, we follow the same fitting procedure as for the SDSS spectrum, except that the relative fluxes of the narrow emission lines are fixed to the SDSS values. The initial fitting reveals significant excess emission on the red side of $\Hb$, which cannot be adequately reproduced by the [\ion{O}{3}] wing alone. Because of the lower spectral resolution, however, the excess emission and the [\ion{O}{3}] wing cannot be reliably separated. We therefore use a single broad Gaussian to jointly account for the [\ion{O}{3}] wing and the red-side excess. As the relative narrow-line fluxes are fixed, the [\ion{O}{3}] core fluxes are well constrained. A similar component is also included on the red side of $\Ha$ to improve the fit.

For the LJT spectra, we further simplify the fitting procedure by restricting the fitting range to 4200--7000~\AA\ and modeling the continuum with a smoothly broken power law and the optical \ion{Fe}{2} emission, while neglecting the host contribution because of the dominant blue continuum.
The relative fluxes of the narrow emission lines are again fixed to the SDSS values. Similar excess emission is seen on the red sides of both $\Hb$ and $\Ha$, as in the HS spectrum, and the $\Hb$ excess cannot be fully reproduced by the [\ion{O}{3}] wing alone. We therefore adopt the same emission-line treatment as for the HS spectrum.

All spectra were flux-calibrated by scaling the modeled [\ion{O}{3}]~$\lambda5007$ core flux to the SDSS value. The fitting results, including the $\Hb$ and $\Ha$ decompositions, are shown in Figure~\ref{fig:spec}, with the broad-line parameters and $\lambda F_{5100}$ listed in Table~\ref{tab:spectra}.
As described above, the narrow-line flux ratios are fixed to the SDSS values across all epochs, with $(\Ha/\Hb)^{\rm na}=5.2$.
The additional broad component in the HS and LJT spectra has FWHM values of $\gtrsim6000$--10,000~km~s$^{-1}$ and fluxes of $(0.4$--$1.1)\times10^{-13}~\cgsF$ in $\Ha$ and $(0.3$--$0.6)\times10^{-13}~\cgsF$ in $\Hb$. Broad \ion{He}{2}~$\lambda4686$ is detected in the HS and LJT spectra but not in the SDSS spectrum. Given the absence of prominent \ion{Fe}{2} features around 4500--4700~\AA\ and 5100--5400~\AA, the broad \ion{He}{2} emission is unlikely to arise from \ion{Fe}{2} contamination.

\begin{deluxetable}{lcccccccccc}
\tabletypesize{\scriptsize}
\tablewidth{0pt}
\tablecolumns{11}
\tablecaption{Broad line and continuum properties of optical spectra\label{tab:spectra}}
\tablehead{
\colhead{epoch} & 
\multicolumn{3}{c}{Flux} & 
\multicolumn{3}{c}{FWHM} & 
\colhead{$\lambda F_{5100}$} & 
\colhead{$\log\MBH$} & 
\colhead{$\log\Lbol$} & 
\colhead{$\log\REdd$} \\
\colhead{(1)} & 
\colhead{$\Ha$} & \colhead{$\Hb$} & \colhead{\ion{He}{2}} & 
\colhead{$\Ha$} & \colhead{$\Hb$} & \colhead{\ion{He}{2}} & 
\colhead{(8)} & \colhead{(9)} & \colhead{(10)} & \colhead{(11)} \\
\colhead{} & 
\colhead{(2)} & \colhead{(3)} & \colhead{(4)} & \colhead{(5)} & 
\colhead{(6)} & \colhead{(7)} & 
\colhead{} & \colhead{} & \colhead{} & \colhead{}
}
\startdata
HS   & 5.08$\pm$0.20 & 1.14$\pm$0.07 & 0.27$\pm$0.04 & 
        3.50$\pm$0.12 & 4.00$\pm$0.51 & 6.24$\pm$0.59 & 
        35.65$\pm$8.31 & 8.49 & 45.31 & -1.29 \\
SDSS & 1.15$\pm$0.01 & 0.16$\pm$0.01 & $<$0.02 & 
        5.40$\pm$0.03 & 5.61$\pm$0.03 & --- & 
        6.37$\pm$0.14 & 8.61 & 44.62 & -2.09 \\
LJ110 & 1.09$\pm$0.04 & 0.45$\pm$0.03 & 0.28$\pm$0.03 & 
        3.77$\pm$0.07 & 4.43$\pm$0.13 & 12.33$\pm$0.60 & 
        52.10$\pm$4.86 & 8.28 & 45.46 & -0.92 \\
LJ207 & 1.00$\pm$0.06 & 0.44$\pm$0.07 & 0.18$\pm$0.03 & 
        3.77$\pm$0.16 & 4.10$\pm$0.17 & 12.00$\pm$0.60 & 
        52.42$\pm$5.01 & 8.26 & 45.46 & -0.90 \\
\enddata
\tablecomments{Columns: (1) observation epoch; (2)–(4) broad line fluxes for $\Ha$, $\Hb$, and \ion{He}{2}, respectively, in $10^{-13}~\cgsF$; (5)–(7) corresponding FWHMs in $10^3$~km~s$^{-1}$; (8) continuum $\lambda F_{\lambda}$ at 5100\,\AA\ in $10^{-13}~\cgsF$; (9) black hole mass in units of $M_{\odot}$; (10) bolometric luminosity in erg~s$^{-1}$; (11) Eddington ratio.}
\end{deluxetable}

\subsubsection{Seyfert Type Transition and Spectral Evolution}
According to the classification scheme of \citet[][Appendix~A1]{Maiolino1995}, the measured $\Hb$/[\ion{O}{3}] ($\simeq$ 4.1, 0.7, and 1.6) classify the HS, SDSS, and LJT spectra as Seyfert Types~1.0, 1.5, and 1.2, respectively. The observed subtype transition occurs on a decadal timescale, similar to the $\Hb$/[\ion{O}{3}] variability reported by \citet{Amrutha2026}, confirming \srcname{} as a recurrent intermediate-type CLAGN.

Across all epochs, the dominant behavior is a strong BWB continuum trend. During the brightest state, the two LJT spectra exhibit consistent luminosity, continuum shape, and broad-line properties over a one-month interval, with the rest-frame 4200~\AA{} power-law flux density nearly $15\times$ that of SDSS.
The earlier HS spectrum exhibits an intermediate continuum level, higher than that of SDSS but lower than those of LJT.

Broad Balmer lines are present in all spectra. The broad $\Ha$ flux drops from $5.08$ to $1.15 \times 10^{-13}~\cgsF$ between HS and SDSS but is comparable between the SDSS and LJT epochs ($1.09 \times 10^{-13}~\cgsF$), despite the much brighter continuum.
In contrast, broad $\Hb$ and $\Hg$ increase significantly from SDSS to LJT.
Consequently, the broad-line Balmer decrement $(\Ha/\Hb)^{\rm br}$ evolves from 4.5 (HS) to 7.2 (SDSS) and then drops sharply to 2.4 (LJT), anti-correlated with the AGN luminosity.

Broad \ion{He}{2}~$\lambda4686$ also shows dramatic behavior (orange lines; Figure~\ref{fig:spec}b--e): it is detected with $\gtrsim 7\sigma$ significance in high states (HS and LJT), but is undetected ($<1\sigma$) in the SDSS spectrum, indicating changing-look behavior in the high-ionization emission.

\subsubsection{Black Hole Parameters}
The bolometric luminosity $\Lbol$ was derived from the 5100~\AA\ continuum using the quasar correction of \citet{Runnoe2012},
\begin{equation}
\log\left(\frac{\Lbol}{\cgsL}\right)
= 4.89 + 0.91\log\left(\frac{\lambda L_{5100}}{\cgsL}\right),
\end{equation}
and the virial black hole mass $\MBH$ from the broad $\Ha$ line using \citet{Shen2011},
\begin{equation}
\begin{split}
\log\left(\frac{\MBH}{\Msun}\right)
= 0.379 + 0.43\log\left(\frac{L_{\Ha}}{10^{42}~\cgsL}\right)
\\
+2.1\log\left(\frac{{\rm FWHM}_{\Ha}}{{\rm km~s^{-1}}}\right).
\end{split}
\end{equation}

Table~\ref{tab:spectra} lists the derived parameters. $\log\MBH$ remains consistent across epochs ($\approx8.5$), while $\log\Lbol$ ranges from $44.62$ to $45.46$ and $\log\REdd$ from $-2.09$ to $-0.90$. Both decline from HS to SDSS and then rise to their highest values in the LJT epoch.

\subsection{X-ray Spectral Analysis}
\label{sec:xray}

\subsubsection{X-ray Flux and Spectral Measurements}
\label{sec:xray_flux}

We analyzed the X-ray spectra (ROSAT, Swift/XRT, and EP/FXT) of \srcname{} for each observation. Spectral fitting was performed in \texttt{XSPEC} (v12.15.0; \citealt{Arnaud1996}) using $C$-statistics \citep{Kaastra2017}. All spectra were grouped to a minimum of two counts per bin, and uncertainties are reported at the 90\% confidence level ($\Delta C = 2.706$). Each spectrum was modeled with a simple absorbed power law (\texttt{phabs}$\times$\texttt{zpowerlw} in \texttt{XSPEC}), with the absorption column density fixed to the Galactic value ($N_{\mathrm{H}} = 3.32\times10^{20}$~cm$^{-2}$; \citealt{HI4PI2016}). 
The 0.5--8.0~keV fluxes $\Fxray$ were derived using the \texttt{cflux} convolution model.
As FXT-A and -B observed our target simultaneously, their spectra were fitted jointly, allowing only a cross-normalization constant between the two modules.

This model provided acceptable fits ($C/\nu = 0.37$--$1.39$) to all XRT spectra except the shortest exposure (0.157 ks, taken during \epochq{}). Owing to low S/N, this exposure yielded a poorly constrained fit ($C/\nu = 9.4/4$) with $\Gamma = 3.2^{+2.0}_{-1.5}$ and $\Fxray = 1.4^{+2.1}_{-0.8}\times10^{-12}~\cgsF$. For the well-fitted epochs, $\Fxray$ ranges from $2.1\times 10^{-12}$ to $8.3\times 10^{-12}~\cgsF$ and the photon indices span $\Gamma = 1.2\pm0.6$ to $2.2\pm0.4$.
The ROSAT and FXT spectra were likewise well described by the same model. The ROSAT spectrum yields $\Fxray = 7.93^{+1.47}_{-1.24} \times 10^{-12}$~erg~cm$^{-2}$~s$^{-1}$ and $\Gamma = 1.38^{+0.25}_{-0.28}$ ($C/\nu = 83.86/103$), while the FXT spectrum gives $\Fxray = 5.47^{+0.43}_{-0.40} \times 10^{-12}$~erg~cm$^{-2}$~s$^{-1}$ and $\Gamma = 1.96\pm0.08$ ($C/\nu = 437.72/433$).

\begin{figure}[hbt!]
\centering
\includegraphics[width=\columnwidth]{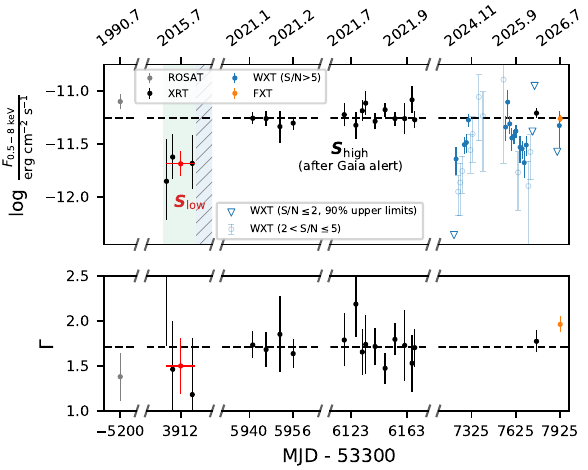}
\caption{Long-term X-ray monitoring of \srcname\ from 1990 to 2026, combining observations from ROSAT (grey point), Swift/XRT (black points), EP/FXT (orange point), and EP/WXT (blue symbols). The red point and black horizontal line show the stacked low- and high-state XRT results. All error bars correspond to the 90\% confidence level.
\textit{Top}: Intrinsic 0.5--8.0 keV flux evolution. WXT bins (bin size = 14~days) are shown as blue filled points ($\mathrm{S/N}>5$), open circles ($2<\mathrm{S/N}\le5$), and downward triangles (90\% upper limits for $\mathrm{S/N}\le2$). The shaded region indicates the \epochq{} interval, as in Figure~\ref{fig:lc}.
\textit{Bottom}: Evolution of the X-ray photon index ($\Gamma$).
\label{fig:xraylc}}
\end{figure}

To improve S/N, we stacked three low-state XRT spectra and all 15 high-state XRT spectra, respectively, using the \texttt{addspec} tool. The stacked low-state spectrum yielded a best-fit $\Gamma=1.50 \pm 0.31$ and $\Fxray = 2.05^{+0.63}_{-0.49}\times 10^{-12}~\cgsF$ ($C/\nu = 22.96/30$). The stacked high-state spectrum was softer, with $\Gamma=1.71 \pm 0.05$ and $\Fxray = (5.58\pm0.24)\times10^{-12}~\cgsF$ ($C/\nu = 370.07/390$). The resulting X-ray LC is shown in Figure~\ref{fig:xraylc}, including the stacked results. The X-ray flux follows the optical evolution, with the post-alert flux significantly exceeding that during \epochq{}.
The ROSAT flux is comparable to the recent bright XRT and FXT measurements, recording an earlier bright X-ray phase in 1990, consistent with a high optical flux state before \epochq{}.

The EP/WXT all-sky survey substantially fills the temporal gaps over the past two years.
We converted the WXT count rates into approximate 0.5--4.0~keV fluxes using an empirical energy conversion factor of $2\times10^{-9}$~erg~cm$^{-2}$~cts$^{-1}$, adopted solely for order-of-magnitude characterization under the current instrumental calibration \citep{Cheng2025, Cheng2026}. Assuming a power-law spectrum with Galactic absorption and a photon index fixed at the stacked high-state XRT value ($\Gamma=1.71$), we estimated the intrinsic 0.5--8.0~keV fluxes and include them as blue symbols in Figure~\ref{fig:xraylc} (with $\mathrm{S/N}\le2$ bins shown as 90\% upper limits).
The estimated fluxes agree well with the nearly contemporaneous FXT observation and are broadly consistent with the XRT measurements, although they are systematically lower than the high-state XRT level. Given the approximate nature of the conversion, we do not investigate this offset further. We also note that, the first WXT bin, despite its low S/N ($<1$), has a substantial exposure of $\sim86$~ks and yields $\Fxray\lesssim4\times10^{-13}$~erg~cm$^{-2}$~s$^{-1}$, indicating that the source was indeed in a relatively faint X-ray state during \epochh{}.

\subsubsection{Spectral Variability and Absorption Constraints}

\begin{figure}[hbt!]
\centering
\includegraphics[width=0.7\columnwidth]{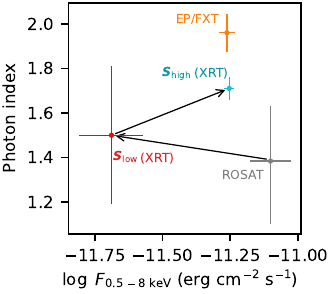}
\caption{
Photon index $\Gamma$ vs.\ 0.5--8~keV X-ray flux, with 90\% error bars.
The stacked low- and high-state Swift/XRT spectra exhibit a tentative SWB trend, further reinforced by the recent EP/FXT observation.
\label{fig:xraycolor}}
\end{figure}

The two stacked Swift/XRT spectra suggest a mild softer-when-brighter (SWB) trend, although their $\Gamma$ uncertainties overlap, with recent EP/FXT observation providing further support for this trend (Figure~\ref{fig:xraycolor}).
The ROSAT best-fit spectrum is harder than the stacked high-state XRT spectrum at a comparable flux. However, the large ROSAT uncertainty in $\Gamma$ and the extrapolation of its flux beyond the PSPC bandpass (0.1-2.4~keV) limit this comparison.

Since the SWB trend can be associated with changes in LoS absorption, we added an intrinsic absorption component (\texttt{zphabs}) to the Galactic-absorbed power-law model and fitted FXT, ROSAT, and the two stacked XRT spectra independently.
All four spectra yielded only 90\% upper limits on the intrinsic absorption column, with $N_{\rm H}^{\rm intr} < 5.10 \times 10^{20}~\mathrm{cm}^{-2}$ (FXT), $N_{\rm H}^{\rm intr} < 5.12 \times 10^{20}~\mathrm{cm}^{-2}$ (ROSAT), $5.06 \times 10^{20}~\mathrm{cm}^{-2}$ (low-state XRT), and $4.38 \times 10^{20}~\mathrm{cm}^{-2}$ (high-state XRT).

As a complementary test, we jointly fitted the two stacked XRT spectra to examine whether variable absorption alone could account for the observed X-ray variability, allowing $N_{\rm H}^{\rm intr}$ to vary independently among epochs.
To avoid the high-state spectrum dominating the fit, we rebinned these spectra to comparable numbers of bins (22–25 each). We first tied both $\Gamma$ and the power-law flux across epochs, obtaining a poor fit of $C/\nu=108.69/43$. We then allowed the power-law flux to vary independently while keeping $\Gamma$ tied, which significantly improved the fit to $C/\nu=44.01/42$, with $\Gamma=1.74^{+0.09}_{-0.07}$.

Overall, the X-ray spectra provide no significant evidence for intrinsic absorption, although a weak absorber ($<10^{21}~\rm cm^{-2}$) cannot be excluded.

\subsection{Multi-wavelength Variability}
\label{sec:variability}
\subsubsection{Overview of the Multi-wavelength Variability}
The multi-wavelength LCs of \srcname{} reveal dramatic variability from X-rays to the MIR (Figures~\ref{fig:lc} and \ref{fig:xraylc}).
In the optical, the source evolved from an early bright state into a prolonged faint phase, remained quiescent for $\sim5$~yr (\epochq{}), and then underwent a rapid rebrightening following the Gaia alert. 
The optical rise associated with the 2020 outburst lasted $\sim1.7$~yr (2019.11--2021.8), during which the source brightened by $\sim1.7$~mag in the $g$ band.
Despite returning to a bright phase, the optical emission continued to vary significantly over the subsequent several years rather than settling into a steady phase.
Noticeable variability is also present during \epochq{} in both the optical and MIR, indicating that the central engine remains active during its faintest phase.
Over the nearly two decades, the source exhibits a peak-to-peak variability amplitude of $\Delta g = 2.29\pm0.03$~mag (a factor of $\sim$8.2 in flux).
Beyond the overall flux changes, the ZTF $gri$ LCs display a clear BWB trend (i.e., $\Delta g > \Delta r > \Delta i$; Figure~\ref{fig:lc}b), with several short-duration optical flares superimposed on the long-term evolution. These behaviors are widely observed in quasar variability (e.g., \citealt{Ruan2014, Guo2016}).

The UV, MIR, and X-ray emissions all broadly track the long-term optical evolution. The UV emission exhibits an even larger peak-to-peak amplitude of $\Delta UVM2 = 2.87\pm0.13$~mag (a factor of $\sim14.1$ in flux). The extreme and highly correlated optical/UV variability establishes \srcname{} as an extreme variability Seyfert galaxy, analogous to the extreme variability quasars (EVQs; \citealt{Rumbaugh2018}).
The Swift/XRT flux increased by $\sim3\times$ between \epochq{} and \epochh{}, while the ROSAT flux measured in 1990 was comparable to the recent bright X-ray measurements. The MIR LC exhibits clear delays during both the declining and rising phases and lacks the short-timescale structures seen in the optical, as expected from dust reprocessing in an extended torus.

In contrast to the strong variability observed at shorter wavelengths, the radio emission remains relatively stable, with a flux density of $\sim$2~mJy throughout the ASKAP monitoring period from 2020 May to 2026 January. 
A $\chi^2$ test against a non-variable hypothesis yields $\chi^2=10.45$ for 11 degrees of freedom ($p=0.49$), suggesting that the observed radio fluctuations are consistent with measurement uncertainties. Notably, these observations cover both sides of the 2020 outburst, yet no corresponding change in the radio flux density is observed.

\subsubsection{Flux Changes Across the 2020 Outburst}
\begin{figure}[hbt!]
\centering
\includegraphics[width=0.9\columnwidth]{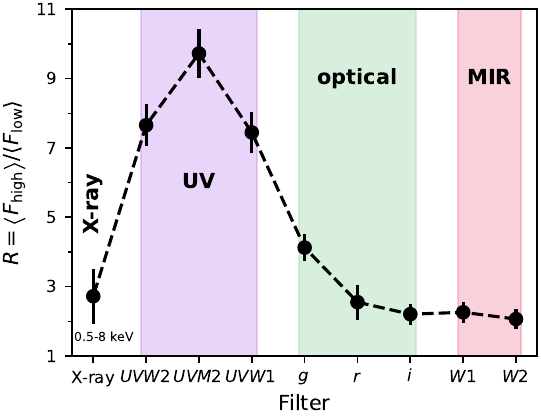}
\caption{
Multi-wavelength flux ratio $R=\Fh/\Fq$ of \srcname{}, derived from WISE ($W1$ and $W2$), ZTF ($g$, $r$, and $i$), Swift/UVOT ($UVW1$, $UVM2$, and $UVW2$), and Swift/XRT observations. Here, $\Fq$ and $\Fh$ represent the characteristic flux levels during \epochq{} and \epochh{}, respectively.
\label{fig:amp}}
\end{figure}

The 2020 outburst marks a clear turning point in the long-term flux variability of \srcname{}. To compare the multi-wavelength fluxes between the pre- and post-outburst epochs, we define the flux ratio $R\equiv\Fh/\Fq$, where $\Fq$ and $\Fh$ are the median fluxes during \epochq{} and \epochh{}, respectively.
Uncertainties are estimated from Monte Carlo simulations that account for photometric errors and finite sampling.
Figure~\ref{fig:amp} shows that $R$ increases systematically from the MIR to the UV.
For the X-ray regime, using the 0.5--8~keV flux ratio from the stacked Swift/XRT spectra (Section~\ref{sec:xray}), we obtain $R_X=2.7\pm0.8$, comparable to or slightly below the $g$-band value $R_g=4.1\pm0.4$ and significantly smaller than the UV amplitudes.

\subsubsection{Variability in the High-state}
\begin{figure}[hbt!]
\centering
\includegraphics[width=\columnwidth]{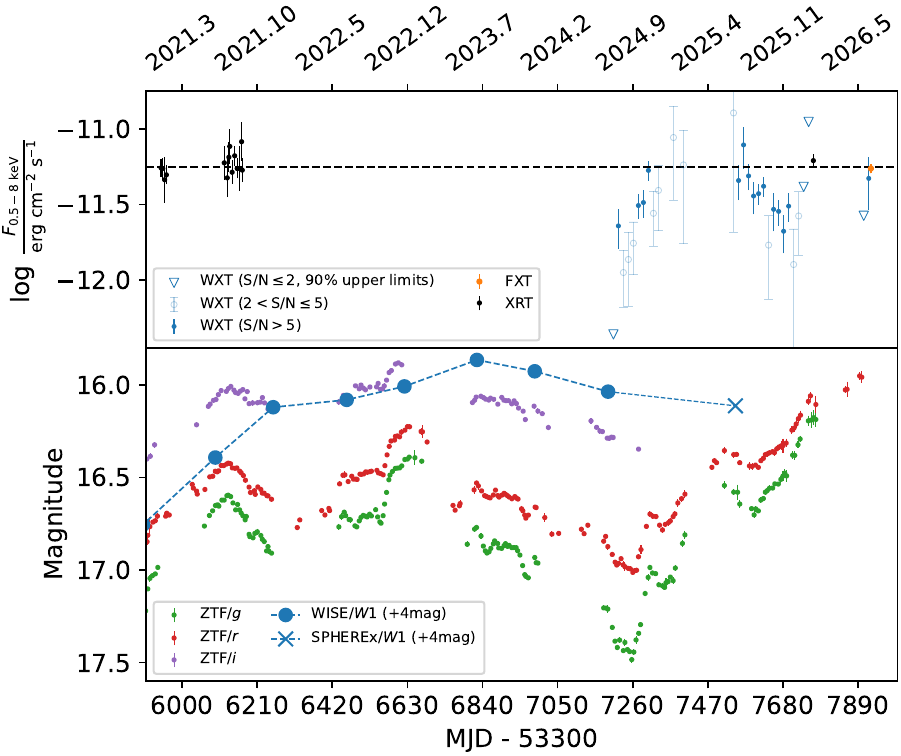}
\caption{Enlarged view of the multi-wavelength LCs during \epochh{}. The symbols are the same as those in Figures~\ref{fig:lc} and \ref{fig:xraylc}. \textit{Top}: X-ray LCs in the 0.5--8 keV band, including Swift/XRT, EP/FXT, and EP/WXT data. \textit{Bottom}: Optical $gri$ LCs from ZTF together with the $W1$-band data from WISE and SPHEREx.
\label{fig:wxtlc}}
\end{figure}

As noted above, the optical emission remained highly variable throughout \epochh{}. The $g$ band exhibited a peak-to-peak variability of $\sim$1.3~mag (a factor of $\sim$3.3 in flux) during \epochh{}.
Figure~\ref{fig:wxtlc} zooms in on \epochh{}, showing the contemporaneous X-ray, optical, and MIR LCs. A pronounced local minimum occurred around 2024 September, after which the optical flux showed an overall rising trend, despite several small month-scale fluctuations.
The MIR LC follows the same overall evolution as the optical emission, but with smoother variations.

The EP/WXT monitoring covers the period during which the optical emission gradually recovered from its local minimum. Notably, the first WXT bin with a faint upper limit was obtained near the optical local minimum. Although the X-ray sampling is much sparser than that of the optical monitoring, the available data suggest that the X-ray evolution broadly follows the optical trend even during \epochh{}.
Unlike the optical LC, the WXT LC exhibits much stronger fluctuations on monthly to sub-year timescales, with the inferred X-ray flux varying by a factor of $\gtrsim18$ within the WXT data alone, based on the ratio between the upper limit of the first WXT bin and the maximum flux among the high-significance detections ($\mathrm{S/N}>5$).
This inferred variability is subject to uncertainties in the flux conversion, including those associated with the current WXT effective-area calibration (systematic uncertainty $\lesssim10\%$; \citealt{Cheng2026}) and the assumed fixed $\Gamma$. However, the relatively stable spectral shape in \epochh{} (Figure~\ref{fig:xraylc}), together with the limited calibration uncertainty, suggests that these effects are unlikely to erase the large-amplitude variability, which remains close to an order of magnitude.

\subsection{Color Variations}
\label{sec:color}
\subsubsection{Bluer-When-Brighter Trend in the Optical}
\label{sec:opt_color}
As noted in Sections~\ref{sec:opt_spectra} and \ref{sec:variability}, \srcname{} exhibits a BWB trend in the optical.
To examine the color behavior in more detail, we rebinned the ZTF $g$ and $r$ photometry onto a common time grid: each $g$-band epoch was paired with the closest $r$-band observation within 10 days; unpaired points were discarded. The resulting $g-r$ vs.\ $g$ color–magnitude diagram (CMD) is shown in Figure~\ref{fig:colors} (left).
The ZTF CMD confirms the BWB trend again. At magnitudes brighter than $g=17.6$~mag, $g-r$ follows a tight linear relation with Pearson $r = 0.93$ ($p = 1.5\times10^{-17}$) and best-fit $g-r = 0.24g-3.87$ (i.e., $\Delta g/\Delta r = 1.32$).
For $g > 17.6$~mag, the colors lie systematically above (i.e., redder than) this linear relation, with a maximum excess of $\Delta(g-r)=0.22\pm0.03$~mag.

\subsubsection{Long-term Mid-infrared Color Variation}
\label{sec:mir_color}
\begin{figure}[hbt!]
\centering
\includegraphics[width=\columnwidth]{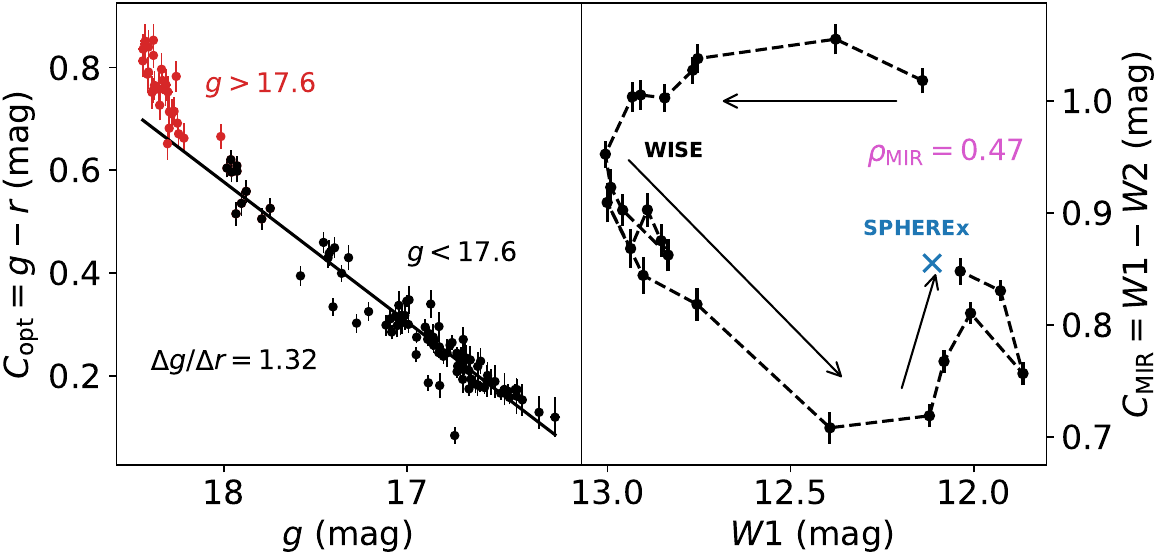}
\caption{
Observed optical and MIR color--magnitude diagrams.
\textit{Left}: ZTF $g-r$ vs. $g$. The reddest points at faint magnitudes (red) lie systematically above the linear relation defined by the brighter data (black).
\textit{Right}: WISE $W1-W2$ vs. $W1$. The dashed line traces the temporal evolution of the source (arrows indicate the direction of evolution), and the blue cross marks the SPHEREx measurement. The source is formally classified as a weak MCV object ($\coefmir=0.47$), while exhibiting a prominent MIR color loop.
\label{fig:colors}}
\end{figure}

We further investigated the long-term MCV of \srcname{} using the MIR LCs. The Pearson correlation coefficient between MIR color $\cmir$ and $W1$ magnitude is $\coefmir=0.47$ ($p = 0.02$, $N=25$), indicating a moderate but statistically significant correlation.
Following the \papers\ classification, \srcname{} falls into the ``weak MCV'' category ($\left|\coefmir\right| < 0.6$).
Additionally, we note that the MIR CMD ($\cmir$ vs. $W1$; Figure~\ref{fig:colors}, right panel) reveals an intriguing loop.
During the initial MIR fading phase, $\cmir$ remained nearly constant; the source then brightened rapidly with a pronounced BWB trend, before appearing to return toward the loop's starting point.

As shown in \papers{}, RWB Seyfert~1s generally exhibit bluer minimum MIR colors ($\cmir{min}$) and larger MCV amplitudes ($\Delta\cmir=\cmir{max}-\cmir{min}$) than BWB Seyfert~1s, with $\cmir{min}=0.7$ marking the approximate boundary between the two populations. Previously known turn-on/off CLAGN predominantly overlap with the RWB population.
For \srcname{}, we measure $(\cmir{min}, \Delta\cmir) = (0.71, 0.35)$. Its $\cmir{min}$ lies near the RWB/BWB boundary, while its large $\Delta\cmir$ is $>5\sigma$ above the BWB population but close to the RWB Seyfert~1 median ($=0.31$). In the $\cmir{min}$--$\Delta\cmir$ plane, it lies within the 80.2\% RWB and 94.4\% BWB highest-density contours, slightly favoring the RWB regime.

Overall, the weak MCV and its intermediate position in the $\cmir{min}$--$\Delta\cmir$ plane indicate that \srcname{} cannot be unambiguously associated with either the BWB or RWB population. Nevertheless, as an intermediate-type CLAGN, its large $\Delta\cmir$ is similar to that of turn-on/off CLAGN. Moreover, most turn-on/off CLAGN (e.g., \citealt{QYang2018, QYang2025}) exhibit an overall negative trend in the $\cmir$--$W1$ diagram, in contrast to the positive $\coefmir$ observed in \srcname{}. We discuss these unusual MCV properties in Section~\ref{sec:dis_mcv}.

\section{Spectral Energy Distribution and Host-Galaxy Decomposition}
If the host contribution is significant in \epochq{}, host contamination would dilute the observed color variations (Figure~\ref{fig:colors}) and suppress the flux ratios $R$ (Figure~\ref{fig:amp}) in the nucleus.
Furthermore, the presence of obscuring material could provide an explanation for the redder optical continuum and higher Balmer decrement in \epochq{}. The main objective of this section is therefore to assess the host contribution and search for evidence of obscuration through spectral energy distribution (SED) analysis.
Given the dramatic variability of \srcname{}, we first examine the evolution of its SED across different epochs (Section~\ref{sec:sed}) and then perform SED decomposition (Section~\ref{sec:sed_analysis}). We finally examine the influence of host on color variations (Section~\ref{sec:subhost_color}) and flux variations (Section~\ref{sec:subhost_amp}).

\subsection{Multi-epoch SEDs}
\label{sec:sed}
\begin{figure}[hbt!]
\centering
\includegraphics[width=0.8\columnwidth]{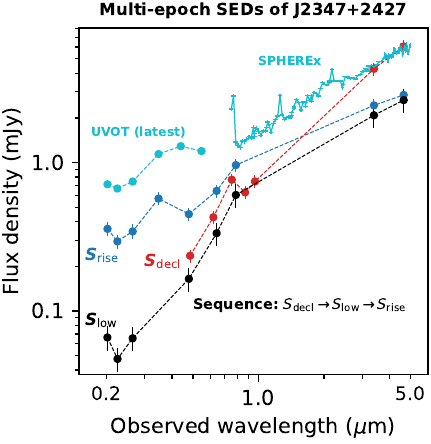}
\caption{
Multi-epoch SEDs of \srcname{}. Average flux densities for \epochq{} (black), \epochd{} (red), and \epochr{} (blue). Cyan symbols denote the most recent Swift/UVOT photometry and SPHEREx spectrum. MIR flux-density variations are not fully synchronized with the optical/UV variability, showing an apparent time delay. \label{fig:sed}}
\end{figure}

To examine the SED evolution, we constructed multi-epoch SEDs by combining photometry from ZTF, PS1, WISE, and Swift/UVOT.
Two 150-day windows, marked by gray shaded regions in Figure~\ref{fig:lc}, sample the declining (\epochd{}) and rising (\epochr{}) phases.
The low state \epochq{} is also included, with UV coverage supplied by three Swift/UVOT observations obtained during the early part of \epochq{} (green shaded region in Figure~\ref{fig:lc}).
For each window, average flux densities were computed in all available bands, and the resulting SEDs are presented in Figure~\ref{fig:sed}, together with the most recent Swift/UVOT photometry and the SPHEREx spectrum (cyan symbols).

The multi-epoch SEDs further illustrate the dust-delayed response: from \epochd{} to \epochq{}, the optical flux decreased only mildly while the MIR flux dropped dramatically; from \epochq{} to \epochr{}, the optical/UV flux rose rapidly, but the MIR brightened more gradually and did not recover to the \epochd{} level.
The most recent Swift/UVOT and SPHEREx data show a further mild increase in the optical/UV continuum. At the same time, the MIR emission now exceeds the \epochd{} flux, consistent with the dust needing sufficient time to respond to the luminosity changes.

\subsection{SED Decomposition for the Low State}
\label{sec:sed_analysis}
Based on the analysis in Section~\ref{sec:sed}, we consider \epochq{} to be the period least affected by dust-heating delays.
We decomposed the \epochq{} photometry using \texttt{CIGALE} (v2025.0; \citealt{Boquien2019}), which enforces energy balance between dust-absorbed stellar light and its re-emission.
The AGN component (disk + torus) was modeled with the \texttt{SKIRTOR} UV-to-IR templates \citep{Stalevski2012, Stalevski2016}.
To account for potential LoS extinction in the SED modeling, we activated the ``polar dust'' module in \texttt{CIGALE}, which acts as a dust screen in front of the accretion disk, without specifying its radial location, and is characterized by various empirical extinction curves \citep{Yang2020}.
The principal parameter ranges are:
\begin{itemize}
\item Star formation history: delayed model with $e$-folding time $\tau_{\rm main} \in [500,\,8000]$\,Myr, age = 5000\,Myr, no burst.
\item Stellar populations \citep{Bruzual2003}: Chabrier IMF \citep{Chabrier2003}, $Z=0.02$.
\item Dust attenuation \citep{Charlot2000}: $A_V^{\rm ISM} \in [0.01,\,1.0]$\,mag, 
\texttt{slope\_ISM} = $-$0.7, \texttt{slope\_BC} = $-$1.3, \texttt{mu} = 0.44.
\item AGN torus: $\Theta \in [20^\circ,\,60^\circ]$, 
$\tau_{9.7} \in \{3,5,9\}$, $p,q \in \{0.0,1.0,1.5\}$, $R_{\rm out}/R_{\rm in}=20$, 
$i \in [0^\circ,\,50^\circ]$.
\item Polar dust: SMC law, $E(B-V) \in [0,\,0.7]$\,mag.
\end{itemize}
We kept all other parameters at the \texttt{CIGALE} defaults.

\begin{figure}[hbt!]
\centering
\includegraphics[width=\columnwidth]{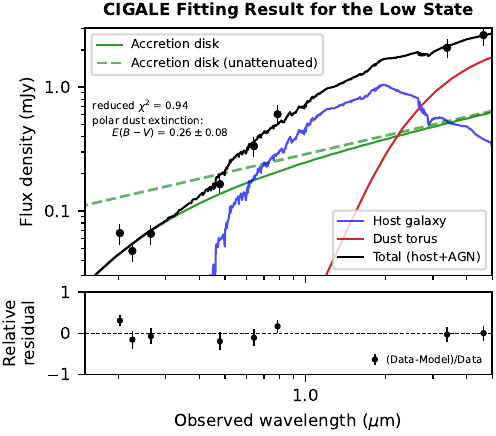}
\caption{
\texttt{CIGALE} SED fitting results for \epochq{}.
Individual model components are shown as solid lines.
The green dashed line indicates the intrinsic (unattenuated) accretion disk SED adopted in \texttt{CIGALE} based on \cite{Feltre2012}.
\label{fig:cigale}}
\end{figure}

The best fit yields a reduced $\chi^2 = 0.94$ (Figure~\ref{fig:cigale}), with a torus half-opening angle $\Theta \approx 37^\circ \pm 10^\circ$ and a low inclination $i \approx 15^\circ \pm 11^\circ$, consistent with the Type~1 nature. The fit requires a moderate polar dust extinction $E(B-V)_{\rm SED} = 0.26 \pm 0.08$~mag, providing evidence for LoS obscuration during the low state. Disabling the polar dust component severely degrades the fit, yielding a reduced $\chi^2$ = 5.5.
During \epochq{}, the host contributes significantly to the observed flux: $27.1\%$ in $g$, $48.3\%$ in $r$, $23.6\%$ in $W1$, and $13.8\%$ in $W2$. Such contamination can significantly dilute the observed nuclear variability, particularly in the optical bands.
The \texttt{BADASS} decomposition of the SDSS/eBOSS spectrum yields host fractions of 25.9\% in $g$ and 36.7\% in $r$.
The $g$-band estimate agrees closely with \texttt{CIGALE}, while the $r$-band fraction is lower, although a direct comparison is limited by differences in aperture coverage ($2''$-diameter) and decomposition methods.

\subsection{Host-subtracted Color Variations}
\label{sec:subhost_color}
\begin{figure}[hbt!]
\centering
\includegraphics[width=\columnwidth]{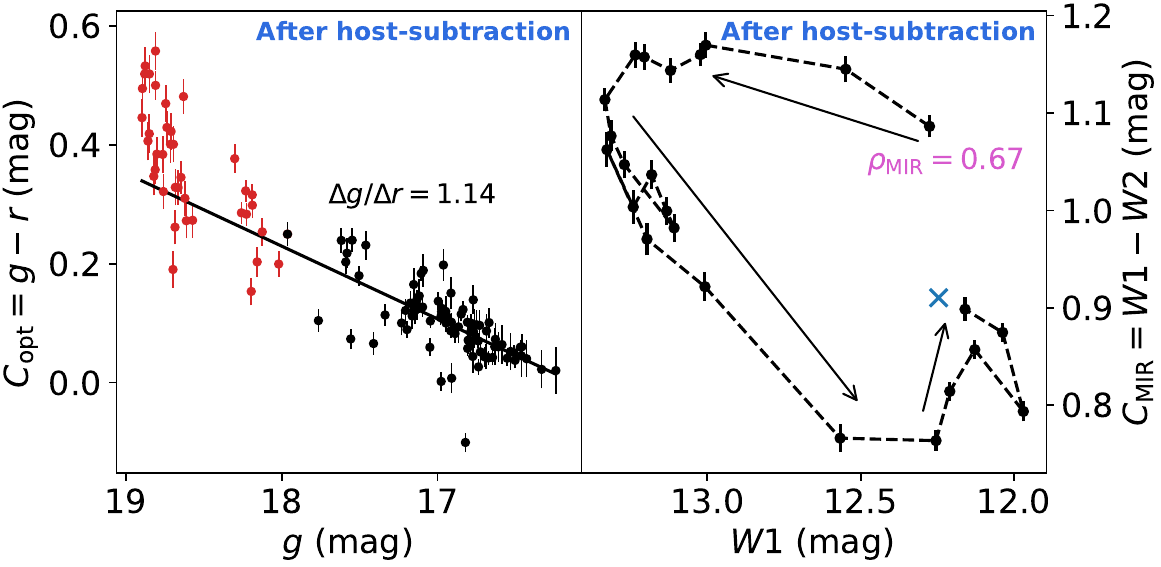}
\caption{Host-subtracted color--magnitude diagrams. The symbols are identical to those in Figure~\ref{fig:colors}.
\textit{Left}: Host-subtracted ZTF $g-r$ vs. $g$. The points at faint magnitudes (red) remain systematically above the linear relation defined by the brighter data (black).
\textit{Right}: Host-subtracted WISE $W1-W2$ vs. $W1$. After host subtraction, the source exhibits a BWB trend ($\coefmir=0.67$), while the MIR color loop persists.
\label{fig:colors_subhost}}
\end{figure}

The SED decomposition (Section~\ref{sec:sed_analysis}) shows that the host contributes appreciable flux in \epochq{}. To isolate the intrinsic nuclear colors, we subtracted the host component and reconstructed the optical and MIR CMDs (Figure~\ref{fig:colors_subhost}).
After host subtraction, the optical colors become bluer ($g-r$ decreases) and the MIR colors redder ($W1-W2$ increases), reflecting the fact that starlight is redder than the disk continuum in the optical but bluer than the hot-dust emission in the MIR.
For the brighter ZTF data, the host-subtracted CMD yields a slope of $\Delta g/\Delta r = 1.14$. The faint points remain systematically offset from the linear relation, with a maximum offset of $\Delta(g-r)=0.24\pm0.03$~mag.
In the MIR, the host-subtracted $\coefmir$ increases to 0.67, moving \srcname{} from the weak MCV ($\left|\coefmir\right| < 0.6$) to the BWB category ($\coefmir > 0.6$). Nevertheless, the prominent MIR color loop persists.

\subsection{Host-subtracted Flux Ratios}
\label{sec:subhost_amp}
\begin{figure}[hbt!]
\centering
\includegraphics[width=0.9\columnwidth]{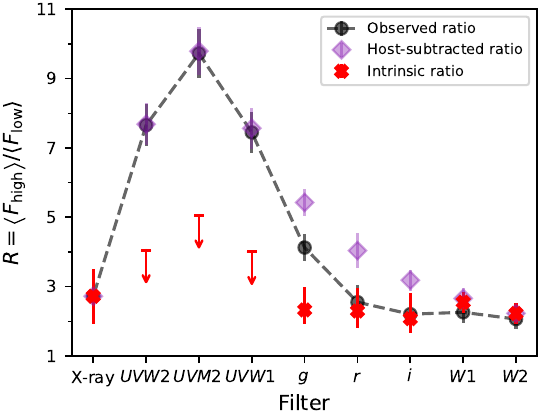}
\caption{
Multi-wavelength flux variation factor of \srcname{}. Black symbols show the observed values (same as Figure~\ref{fig:amp}), purple symbols show the host-corrected values, and red symbols show the intrinsic values after further correcting for continuum extinction.
\label{fig:ampcorr}}
\end{figure}

A non-negligible host contribution dilutes the observed optical and MIR variability. After subtracting the host, the peak-to-peak $g$-band variability factor increases from $\sim8.2\times$ to $12\times$, consistent with the large continuum variation between the SDSS and LJT spectra derived from the spectral decomposition.
We further correct the flux ratios $R$ between the characteristic flux levels during \epochq{} and \epochh{} (purple diamonds in Figure~\ref{fig:ampcorr}). The $g$-band flux ratio increases to $R^{\mathrm{sub\text{-}host}}_g=5.4\pm0.4$, making the contrast with $R_X$ even more pronounced.

\section{Discussion\label{sec:discuss}}
\subsection{A Recurrent Intermediate-type CLAGN Driven by Intrinsic Accretion-State Changes}
\label{sec:clagn}
The optical spectra of \srcname\ over the past three decades reveal a Type~1.0$\rightarrow$1.5$\rightarrow$1.2 transition (Section~\ref{sec:opt_spectra}): the broad Balmer lines weakened from the earliest HS spectrum to the SDSS spectrum and re-strengthened in the recent LJT spectra following the 2020 outburst. This behavior is characteristic of the broader CLAGN population. In tandem with the Balmer-line changes, the X-ray-to-MIR fluxes exhibit coordinated dramatic recurrent variability.

The dramatic state transitions naturally raise the question of whether the observed variability is driven primarily by changes in the LoS obscuration or by intrinsic changes in the accretion flow. Several observational characteristics favor the latter.
The optical LCs exhibit short-timescale stochastic flares superposed on the long-term brightening trend, together with a clear BWB behavior, both of which are characteristic signatures of thermal fluctuations in an inhomogeneous accretion disk \citep{Dexter2011, Cai2016, Cai2018}.
More importantly, the MIR emission varies coherently with the optical/UV continuum and exhibits a large amplitude ($\Delta W1=1.14$~mag). Explaining such a large MIR brightening solely by a reduction in obscuration would require an extreme extinction of $A_V\simeq15$~mag (assuming an SMC extinction law; $R_V^{\rm SMC} = 2.7$; \citealt{Prevot1984}), which is inconsistent with the optical data.
Furthermore, the joint modeling of stacked Swift/XRT spectra requires intrinsic X-ray luminosity variations (Section~\ref{sec:xray}). At the same time, the X-ray spectra provide no compelling evidence for variable gas absorption, yielding only a relatively low upper limit on the absorbing column density.

Although some CLAGN have been suggested to be triggered by TDEs, this scenario appears unlikely for \srcname. The optical LC remained in a high state following the 2020 outburst rather than exhibiting the monotonic $t^{-5/3}$ decline expected for a TDE \citep{Rees1988}.
Moreover, \srcname\ has undergone recurrent changing-look transitions over nearly three decades, making it difficult to reconcile the observed behavior with a single stellar disruption.
In the MIR, \citet{Yao2025} showed that TDEs typically exhibit much more rapid color evolution than CLAGN, with $\cmir$ variation rates (CVRs) generally $> 0.2~{\rm mag~yr^{-1}}$. By comparison, the 2020 MIR outburst of \srcname\ has ${\rm CVR}\approx0.06~{\rm mag~yr^{-1}}$, substantially slower than expected for TDEs.

Recently, \citet{Jana2025} found that spectroscopic changing-look transitions are universally accompanied by substantial changes in Eddington ratio and identified a characteristic transition value of $\log\REdd\sim-2$ for Type~1.0~$\leftrightarrow$~1.8/1.9/2 transitions. The evolution of \srcname\ is broadly consistent with this picture: its Eddington ratio increased from $\log\REdd=-2.09$ to $-0.90$, crossing the proposed transition regime. In particular, the low-state Eddington ratio lies close to the proposed transition value, where the inner optically thick accretion disk may become truncated and replaced by a hot radiatively inefficient flow.
Such a structural change could naturally reduce the contribution from the hottest part of the accretion disk, producing additional reddening that contributes to the offset of the low-state data from the linear optical color--magnitude relation (red points in Figures~\ref{fig:colors} and \ref{fig:colors_subhost}).

Nevertheless, the substantial evolution of the broad-line Balmer decrement across different epochs may also reflect a contribution from variable obscuration (Section~\ref{sec:opt_spectra}). \citet{Dong2008} found that $(\Ha/\Hb)^{\rm br}$ in unobscured AGN are clustered around 3. The much larger value of $7.2$ observed during \epochq{} may therefore indicate additional dust reddening. We further discuss this possibility in Section~\ref{sec:dis_ivoc}.

The radio emission is a notable exception to the coordinated variability: it remained steady at $\sim$2~mJy across the 2020 outburst (Section~\ref{sec:variability}). This may indicate that the radio emission arises from an extended, optically thin region (e.g., shocked gas at the terminus of a jet or outflow) that does not respond promptly to changes in the central engine, and/or that the jet power is governed primarily by the black hole spin rather than by the instantaneous accretion rate.

\subsection{Long-term MCV of the Changing-look Transition}
\label{sec:dis_mcv}
Our interest in \srcname\ was initially motivated by its unusually large MIR variability amplitude and distinctive MCV, as introduced in Section~\ref{sec:intro} (also see the details in Section~\ref{sec:color}). Such a large amplitude is naturally expected for an AGN undergoing an accretion-state transition. Here we focus on the MCV behavior of \srcname{}.

\srcname{} is initially classified as a weak MCV source ($\left|\coefmir\right|=0.47<0.6$). Within the framework of \papers{}, this indicates that the variable nucleus does not fully dominate the MIR emission and that host contamination contributes to the weak MCV, as seen in \epochq{}. After subtracting the host contribution, the MCV strengthens to $\coefmir=0.67$, only slightly above the BWB threshold, but still supports the proposed host-dilution scenario.

In the $\cmir{min}$--$\Delta\cmir$ plane, turn-on/off CLAGN predominantly overlap with the RWB population (\paperI). This can be naturally understood as a consequence of the dramatic transition between the AGN-dominated and host-dominated states. For example, \citet{Sheng2020} showed that turn-off CLAGN evolve from AGN-like MIR colors ($\cmir\gtrsim0.8$) to galaxy-like colors ($\cmir\lesssim0.5$), resulting in both bluer $\cmir{min}$, measured during the host-dominated state, and larger $\Delta\cmir$.

In contrast, \srcname\ does not enter a host-dominated MIR state. Although the host contributes non-negligibly in \epochq{}, it accounts for only $23.6\%$ of the $W1$ emission according to the \texttt{CIGALE} result. Combined with the fact that its broad Balmer lines never completely vanish, this indicates that the nucleus remains active throughout the changing-look cycle.
Thus, its $\cmir{min}$ is instead measured during an active nuclear state, when the surrounding dust is hotter.
This leads to a more moderate MIR color evolution ($\cmir = 0.71$--$1.06$, remaining close to AGN-like color) compared to extreme CLAGN, while still exceeding that of ordinary Seyfert~1s, placing \srcname\ between the RWB and BWB populations in the $\cmir{min}$--$\Delta\cmir$ plane.
The persistent nuclear activity also explains the positive $\coefmir$ of \srcname{}, as its MIR variability is less affected by host dilution than in host-dominated systems, where negative trends are more common, including RWB Seyfert~1s (\papers) and turn-on/off CLAGN.

Intermediate-type CLAGN may similarly retain substantial nuclear activity in their low states without becoming host-dominated, and may therefore exhibit MCV properties more similar to \srcname{} than to turn-on/off CLAGN.

\subsection{MIR Color--Magnitude Hysteresis}
Another remarkable property of \srcname{} is its MIR color--magnitude loop (Figures~\ref{fig:colors} and \ref{fig:colors_subhost}, right panels), which a single-temperature, synchronous dust model in \papers{} cannot reproduce. This loop contributes substantially to the weak MCV, even after host subtraction.

One possibility is a change in the dust distribution, such as widespread grain sublimation during the brightening phase. To test this, we estimated the dust temperature $T_d$ for each WISE epoch by fitting the host-subtracted $W1$ and $W2$ flux densities with a modified blackbody $F_\lambda \propto Q_{\rm abs}(\lambda;a)\,B_\lambda(T_d)$,
where absorption efficiencies $Q_{\rm abs}$ are taken from \citet{1993ApJ...402..441L}.
For typically $a=0.1~\mu$m silicate, silicon carbide, and graphite grains, we obtain temperatures in the ranges 705--928~K, 882--1287~K, and 575--711~K, respectively, all well below sublimation thresholds. Extensive dust sublimation is therefore unlikely to account for the loop.

An extended dust torus with a radial temperature gradient naturally explains the loop. Since dust at smaller radii responds more quickly to changes in the central engine, a brightening event heats the innermost dust first, shifting the MIR color blueward while the cooler outer dust has yet to respond. Conversely, when the central luminosity declines, the inner dust cools promptly and the color reddens, whereas the outer dust still retains its previous elevated temperature. Consequently, at a given observed $W1$ flux, the source appears bluer when it is brightening and redder when it is fading, tracing a loop in the CMD.
Indeed, a MIR hysteresis loop has also been observed in the nucleus of UGC~11487 following a TDE, likely arising from a similar multi-zone dust response \citep{Izviekova2026}.

For \srcname{}, the recurrent changing-look behavior provides particularly favorable conditions for revealing the MIR loop. The circumnuclear dust has undergone repeated cooling and reheating over several years, allowing the delayed response of dust at different radii to become visible in the CMD. A single changing-look episode might reveal only part of the loop, whereas shorter-timescale stochastic AGN variability could partially wash out this hysteresis signature. More extreme changing-look events or other outbursts (e.g., TDEs) may additionally involve substantial dust sublimation, producing a more complex morphology.

\subsection{Broad-line Balmer Decrement and Possible Contribution from Variable Obscuration}
\label{sec:dis_ivoc}
As discussed in Section~\ref{sec:clagn}, the dramatic variability of \srcname\ is primarily driven by an accretion-state transition, potentially accompanied by intrinsic continuum reddening in \epochq{} (e.g., a truncated accretion disk). Nevertheless, we cannot exclude a secondary contribution from variable obscuration.
The substantial variation in the broad-line Balmer decrement provides one line of evidence for variable obscuration. At the same time, dust obscuration could also produce continuum reddening similar to that caused by intrinsic changes in the accretion flow.
Consistent with this interpretation, the SED decomposition of \epochq{} requires a polar dust component with $E(B-V)_{\rm SED}=0.26\pm0.08$ mag.
In contrast, \epochh{} exhibits a very blue continuum, consistent with the removal of the obscurer.
The obscuration during \epochq{} may have arisen from a dusty cloud in the clumpy torus crossing the LoS occasionally, or from dusty material on sub-pc scales that was lifted into the LoS by radiation-driven flows and/or formed in cooled BLR gas \citep[e.g.,][]{Mrk590}.
For an obscurer located on sub-pc scales, dust sublimation could remove it in the 2020 outburst.
Radiation pressure may have further aided the clearing by displacing dusty material out of the LoS \citep{Ricci2017}.

We quantify the dust extinction from the host-subtracted ZTF $g-r$ color.
The maximum deviation from the bright-state relation reported in Section~\ref{sec:subhost_color} corresponds to a dust color excess $E(g-r)_{\rm ZTF} = \Delta(g-r) = 0.24\pm0.03$~mag.
Adopting an SMC extinction law with $k_g - k_r = 1.10$ gives
\begin{equation}
E(B-V)_{\rm ZTF} = \frac{E(g-r)_{\rm ZTF}}{k_g - k_r} = 0.22\pm0.03~\mathrm{mag},
\end{equation}
consistent with the SED-derived value $E(B-V)_{\rm SED}$. We adopt the average $E(B-V)_{\rm cont} = 0.24\pm0.08$~mag as the extinction toward the accretion disk continuum in \epochq{}, corresponding to $A_V^{\rm cont} = R_V^{\rm SMC} \cdot E(B-V)_{\rm cont} = 0.66\pm0.23$~mag.
This value is consistent with the median $A_V=0.60^{+0.32}_{-0.19}$~mag reported for reddened Type~1 AGN by \citet{Hernandez2026}, suggesting a moderate and reasonable level of extinction for a Type~1 AGN.
The extinction estimates above implicitly assume that the observed reddening of the continuum is entirely caused by dust extinction.
Given that the changing-look phenomenon in \srcname\ is primarily driven by intrinsic accretion-state changes, the extinction value derived here should therefore be regarded as an upper limit.

The broad-line Balmer decrement provides an additional probe of the obscuration toward the BLR, evolving from $(\Ha/\Hb)^{\rm br}\simeq4.5$ in the earliest HS spectrum to $7.2$ during the SDSS epoch and $2.4$ in the recent LJT spectra.
If interpreted solely as dust extinction, reproducing the SDSS decrement from the LJT value requires $A_V^{\rm BLR}\approx2.95$~mag, implying a dereddened broad $\Ha$ flux $\sim9\times$ higher than that measured in the LJT spectra. Such a large intrinsic flux in the low state is difficult to reconcile with the observed accretion-state evolution.

In reality, the broad-line Balmer decrement is a complex diagnostic that can be affected by factors beyond dust extinction. The intrinsic $(\Ha/\Hb)^{\rm br}$ depends on the physical conditions of the dense BLR, with radiative-transfer effects and changes in gas density and ionization state potentially producing substantial variations.
Photoionization calculations by \citet{Yan2026} show that, at high BLR densities ($n_{\rm H} \gtrsim 10^9~\mathrm{cm}^{-3}$), the intrinsic $(\Ha/\Hb)^{\rm br}$ can exceed 10 at lower densities and decreases with increasing gas density.
Observationally, \citet{Dong2008} found $(\Ha/\Hb)^{\rm br}$ of 2.3--4.2 in blue AGNs, spanning both below and above the canonical Case B value of $\approx2.86$.
More recently, multi-epoch spectroscopy by \citet{Son2025} revealed systematic variations in the Balmer decrement with the AGN continuum luminosity that cannot be fully explained by standard dust-extinction curves, suggesting that optical-depth effects, collisional excitation in the high-density BLR may play important roles. Similarly, \citet{Chen2026arXiv} found that the elevated broad-line decrements in little red dots are primarily associated with high-density gas rather than dust extinction.
The substantial change in $(\Ha/\Hb)^{\rm br}$ for \srcname\ therefore likely includes a contribution from intrinsic BLR evolution.

A delayed response of the BLR to the changing ionizing continuum may provide another contribution. The ``changing-look sequence'' proposed by \citet{Guo2020} describes a progressive response of the broad emission lines to changes in the ionizing continuum, with $\Hb$ responding more rapidly than $\Ha$. Such a sequence could naturally produce a temporarily enhanced $(\Ha/\Hb)^{\rm br}$ during the fading phase and a lower ratio as the BLR re-equilibrates following the subsequent brightening. This scenario is qualitatively consistent with the observed evolution of the broad-line decrement in \srcname.
This effect may be relevant to the 1997 HS spectrum, which exhibits relatively strong broad $\Ha$ emission despite its intermediate continuum level.
The earlier ROSAT observation in 1990 revealed an X-ray flux comparable to recent measurements, suggesting that \srcname{} has experienced a previous high-activity state.
If the ionizing continuum subsequently declined before the HS observation, the delayed BLR response could have maintained a relatively high broad $\Ha$ flux.

Other factors, such as obscurer kinematics, may also affect the observed decrement. An obscuring structure moving relative to the BLR can modify the observed decrement and produce a large $(\Ha/\Hb)^{\rm br}$ with less extinction than expected for a static foreground screen \citep{Sitaram2026}. In the present case, however, reconciling the SDSS Balmer decrement with the much smaller continuum extinction of $A_V^{\rm cont}$ would require a relative velocity of $\gtrsim1.5\times10^5$~km~s$^{-1}$, which is implausibly large. Thus, the motion of the obscuring material is unlikely to be a significant factor.

\subsection{Relatively Modest X-ray Variability across the Changing-look Transition}
\label{sec:dis_xrayvar}

The X-ray and optical emission exhibit broadly consistent long-term behavior, suggesting that the two bands remain coupled across the changing-look transition.
However, as shown in Figure~\ref{fig:amp}, the X-ray variability ($R_X=2.7\pm0.8$) is smaller than the $g$-band variability, and this difference becomes even more pronounced after subtracting the host contribution (Figure~\ref{fig:ampcorr}, $R^{\mathrm{sub\text{-}host}}_g=5.4\pm0.4$).

The presence of obscuration in \epochq{} may help explain the relatively modest X-ray variability.
We therefore correct the host-subtracted flux ratios for the extinction in \epochq{} derived in Section~\ref{sec:dis_ivoc}.
Because the UV correction is highly sensitive to the adopted extinction curve, we conservatively apply the $g$-band extinction correction to the UV bands, such that the resulting UV flux ratios should be regarded as upper limits.
After correcting for both host dilution and extinction (red crosses in Figure~\ref{fig:ampcorr}), the extinction-corrected optical/UV ratios become comparable to the observed $R_X$.

Nevertheless, the relatively modest X-ray variability may also be intrinsic. If the changing-look transition is primarily driven by long-term variations in the accretion flow, the optical/UV emission from the accretion disk is expected to respond directly to the changing accretion power, whereas the X-ray corona may exhibit a weaker response.

In contrast to its relatively modest long-term variability, EP/WXT monitoring reveals much stronger X-ray fluctuations on sub-year timescales, substantially exceeding those of the contemporaneous optical emission.
Such behavior is consistent with X-ray emission originating from the compact inner corona, whereas the optical emission is produced over the much larger accretion disk.

\subsection{Narrow-line Balmer Decrement}
The narrow-line Balmer decrement, $(\Ha/\Hb)^{\rm na}=5.2$, provides an additional constraint on the dust distribution.
Unlike the BLR, the NLR is a relatively low-density environment where the extreme high-density effects discussed above are not expected to dominate. The deviation $(\Ha/\Hb)^{\rm na}$ from the Case B value can therefore be more naturally attributed to dust extinction. Assuming an intrinsic ratio of $2.86$, this corresponds to $A_V^{\rm NLR}\approx1.6$~mag.

Given the much larger spatial extent of the NLR, if its extinction were produced by an isotropic obscuring screen, the nuclear continuum and broad-line emission would be expected to experience comparable or stronger extinction than $A_V^{\rm NLR}$. Interestingly, the inferred continuum extinction is only $A_V^{\rm cont}\approx0.66$~mag even in the reddest low state, and $(\Ha/\Hb)^{\rm na}$ is also higher than $(\Ha/\Hb)^{\rm br}$ in high states.
This suggests that the dust distribution across the extended NLR is spatially inhomogeneous and anisotropic.

\subsection{\ion{He}{2} $\lambda4686$ Changing-look}
Broad \ion{He}{2} $\lambda4686$ in \srcname\ exhibits a clear changing-look behavior (Section~\ref{sec:opt_spectra}). One possible explanation is variable obscuration. Because \ion{He}{2} originates from the innermost, high-ionization BLR, its emission region should experience dust extinction comparable to that affecting the optical continuum. Under this assumption, the expected low-state \ion{He}{2} flux ($\sim0.1\times10^{-13}~\cgsF$) remains well above the observational upper limit. Therefore, obscuration alone cannot account for the disappearance of broad \ion{He}{2}, implying that intrinsic changes in the ionizing source and/or BLR are required.

Broad \ion{He}{2} requires photons above the He$^+$ ionization threshold and is therefore a sensitive tracer of the EUV ionizing continuum. Its disappearance in the SDSS spectrum indicates that the EUV continuum was substantially suppressed during \epochq{}. Such behavior is expected if the inner accretion flow underwent intrinsic structural changes, for example through a truncated accretion disk at low Eddington ratio discussed in Section~\ref{sec:clagn}. Consequently, the \ion{He}{2} changing-look behavior suggests that the low-state continuum is intrinsically redder, consistent with the multi-epoch SEDs, although a contribution from variable obscuration cannot be fully excluded.

This result is consistent with the \ion{He}{2} and $\Hb$ changing-look quasars studied by \citet{Lu2025}, who found that both classes exhibit large-amplitude optical variability and that broad \ion{He}{2} generally shows stronger variability than the Balmer lines, a trend also seen in \srcname.
Such behavior is naturally understood within the changing-look sequence discussed above. Owing to its significantly higher ionization potential, broad \ion{He}{2} responds more rapidly to changes in the ionizing continuum than the Balmer lines. Consequently, the highest-ionization region of the BLR appears to respond first to both the fading and renewed ionizing continuum, turning off before and recovering ahead of the Balmer-emitting region.

\subsection{Comparison with Extreme Variability Quasars}
With its large-amplitude optical/UV variability on a timescale of only a few years, \srcname\ exhibits EVQ-like behavior despite its Seyfert luminosity. Recent statistical studies of EVQs have shown that they exhibit systematically stronger very broad emission-line components (VBCs) than ordinary quasars, suggesting enhanced activity in the innermost BLR (\citealt{Ren2022}). This behavior is consistent with a two-component BLR consisting of an intermediate line region and a very broad line region located much closer to the SMBH (e.g., \citealt{Wills1993, Brotherton1994, Sulentic2000ARA&A}).

Our multi-epoch spectral decomposition requires an additional very broad Gaussian ($\mathrm{FWHM} \gtrsim 6000$--$10{,}000\ \mathrm{km\,s^{-1}}$) to reproduce the extended line wings in both $\Hb$ and $\Ha$ regions. Although broad wings of [\ion{O}{3}] are commonly observed in the $\Hb$ region and can contribute to the profile asymmetry, their typical widths are substantially smaller than those required here and cannot account for the additional component detected in both $\Hb$ and $\Ha$.
The VBC varies coherently with the optical continuum and \ion{He}{2}. Notably, the \ion{He}{2} line turned off during the low state and re-emerges in the high state, closely mirroring the VBC behavior. This tight coupling implies that the VBC originates in the innermost BLR, making it highly sensitive to variations in the ionizing continuum.

The coexistence of pronounced optical variability, an intermediate-type changing-look transition, and a strongly variable VBC in \srcname\ is broadly consistent with the emerging view that many CLAGN represent a selection-biased subpopulation of extreme variability objects rather than a fundamentally distinct class \citep{Ren2024}.

\section{Conclusions\label{sec:conclusion}}
We present a comprehensive multi-wavelength study of the extreme variability Seyfert galaxy \srcname{}, combining three decades of optical spectroscopy with optical/UV, MIR, radio, and X-ray monitoring.
The source exhibits recurrent intermediate-type changing-look behavior, with Seyfert-type transitions from Type~1.0 to 1.5 and then to 1.2, accompanied by the disappearance and reappearance of broad \ion{He}{2}~$\lambda4686$ and strong evolution of the broad-line Balmer decrement. Long-term monitoring reveals coordinated variability from the optical/UV to the X-ray and MIR, including a loop-like MIR color variation (MCV) and no significant intrinsic X-ray absorption from X-ray spectral analysis.
Our main conclusions are summarized as follows.
\begin{enumerate}

\item
\srcname{} is identified as a recurrent intermediate-type CLAGN, exhibiting repeated Seyfert-type transitions (Type~1.0$\rightarrow$1.5$\rightarrow$1.2) together with coordinated variability from the X-ray to the MIR. The broad \ion{He}{2} line disappears during the low state and re-emerges in the high state, indicating dramatic changes in the ionizing continuum.

\item
The coordinated multi-wavelength variability, the evolution of the Eddington ratio across the proposed transition regime, and the lack of compelling evidence for variable X-ray absorption favor intrinsic changes in the accretion state as the primary driver of the changing-look behavior. Nevertheless, we cannot completely exclude a secondary contribution from variable obscuration.

\item
Host-galaxy subtraction moves \srcname{} from the weak MCV to the bluer-when-brighter (BWB) regime, providing direct support for the host-dilution scenario proposed in \papers{}. Compared with classical turn-on/off CLAGN, \srcname{} exhibits a more moderate MCV, occupying an intermediate position between the BWB and redder-when-brighter (RWB) populations while retaining AGN-like MIR colors throughout the changing-look cycle, consistent with the more moderate state transition of this recurrent intermediate-type CLAGN.

\item
The prominent MIR color loop reveals a delayed response of the torus to the varying accretion disk, indicating that the MCVs cannot be attributed solely to the host-dilution effect proposed in \papers{}, particularly for sources exhibiting non-linear color--magnitude evolution (i.e., weak/no MCV).

\end{enumerate}

With its extreme variability and repeated state transitions, \srcname{} is an excellent laboratory for future time-domain studies, including spectroscopic reverberation mapping campaigns to resolve the BLR structure and trace the accretion--obscuration interplay.

\begin{acknowledgments}
This work was supported by the National Science Foundation of China (NSFC-12573110, 12333002, 12233001, 12221003, 12133001, 11833007 and 12433004) and the China Manned Space Program (CMS-CSST-2025-A07, CMS-CSST-2025-A09).
YLA was supported by the Natural Science Foundation of Top Talent of SZTU (GDRC202208) and the Shenzhen Science and Technology program (JCYJ20230807113910021). LMD also acknowledges the support from the Key Laboratory for Astronomical Observation and Technology of Guangzhou, the Astronomy Science and Technology Research Laboratory of the education department of Guangdong Province.
This work is based on data obtained by Einstein Probe (also known as ``Tianguan'' in Chinese), a space mission supported by the Strategic Priority Program on Space Science of Chinese Academy of Sciences, in collaboration with ESA, MPE, and CNES (Grant No.~XDA15310000, No.~XDA15052100).
This research makes use of data products from the {\it Wide-field Infrared Survey Explorer (WISE)} and the {\it Near-Earth Object Wide-field Infrared Survey Explorer (NEOWISE)}. {\it WISE} is a joint project of the University of California, Los Angeles, and the Jet Propulsion Laboratory/California Institute of Technology; {\it NEOWISE} is a project of the Jet Propulsion Laboratory/California Institute of Technology. {\it WISE} and {\it NEOWISE} are funded by the National Aeronautics and Space Administration. 
This scientific work uses data obtained from Inyarrimanha Ilgari Bundara / the Murchison Radio-astronomy Observatory. We acknowledge the Wajarri Yamaji People as the Traditional Owners and Native Title Holders of the Observatory site. CSIRO’s ASKAP radio telescope is part of the Australia Telescope National Facility (\url{https://ror.org/05qajvd42}). Operation of ASKAP is funded by the Australian Government with support from the National Collaborative Research Infrastructure Strategy. ASKAP uses the resources of the Pawsey Supercomputing Research Centre. Establishment of ASKAP, Inyarrimanha Ilgari Bundara, the CSIRO Murchison Radio-astronomy Observatory and the Pawsey Supercomputing Research Centre are initiatives of the Australian Government, with support from the Government of Western Australia and the Science and Industry Endowment Fund. This paper includes archived data obtained through the CSIRO ASKAP Science Data Archive, CASDA (\url{https://data.csiro.au}).
DeepSeek was used to assist with language editing and polishing of the manuscript. All AI-assisted text was subsequently reviewed, verified, and edited by the authors.
\end{acknowledgments}

\bibliography{sample701}{}
\bibliographystyle{aasjournalv7}
\end{document}